\documentclass{aa}
\usepackage{graphicx}
\usepackage[varg]{txfonts}
\usepackage{pdflscape}
\usepackage{amsmath}
\usepackage{empheq}
\usepackage{nccmath}
\usepackage{gensymb}
\usepackage{amssymb}
\usepackage{textcomp}
\usepackage{xfrac}
\usepackage{xcolor}
\usepackage[colorlinks=true, citecolor=blue, linkcolor=blue, urlcolor=blue]{hyperref}
\usepackage{natbib} 

\begin{document}

\title{Deriving physical parameters of unresolved star clusters}
\subtitle{VIII. Limits of aperture photometry for star cluster studies}
\author{Karolis Daugevi{\v c}ius \and Eimantas Kri{\v s}{\v c}i{\= u}nas \and Erikas Cic{\.e}nas \and Rima Stonkut{\.e} \and Vladas Vansevi{\v c}ius}
\institute{Center for Physical Sciences and Technology, Saul\.{e}tekio av. 3, 10257 Vilnius, Lithuania \\\email{vladas.vansevicius@ftmc.lt}}

\date{Received 21 February 2024; published 13 August 2024}

\abstract
{Recently, it has been noticed that the discrepancies in the integrated colour indices (CIs) between star clusters and models are mostly due to the projection of bright stars in the apertures. In order to reduce this problem, the method of adaptive aperture photometry has been proposed. This method has been applied to star clusters from the \object{M\,31} Panchromatic $Hubble$ Andromeda Treasury (PHAT) survey, and studies show that the adaptive aperture photometry performs better than the conventional approach.}
{The aim of this study is to determine the best achievable limits on the accuracy and applicability of the aperture photometry method for studying star clusters in the local Universe.}
{We computed a large network of artificial 3D star clusters spanning the parameter space of the \object{M\,31} clusters. We then simulated images of these clusters by projecting each onto a 2D plane from 100 directions. Star cluster images were generated in six passbands to match the PHAT survey. To investigate the limiting accuracy of aperture photometry and the limits of its applicability to star cluster studies, we measured the simulated images and performed parameter determination tests.}
{We demonstrate that star clusters with and without post-main-sequence stars have significant photometric differences. We show that in order to obtain reliable physical parameters of star clusters, the CIs must be measured using an aperture with a radius larger than the cluster’s half-light radius. Furthermore, we demonstrate that the parameter determination of young clusters ($\sim$10\,Myr) is problematic regardless of the aperture size used. Therefore, it is advisable to determine the parameters of these clusters using colour-magnitude diagram fitting methods, when possible. We also show that the randomness of the viewing angle can lead to a CI uncertainty of up to 0.1\,mag, depending on cluster parameters and aperture size.} 
{}


\maketitle


\section{Introduction}
   \label{Sec1}

Understanding star formation processes is one of the crucial current astrophysical problems as it is the main galaxy-building mechanism throughout the majority of cosmic time. It is widely accepted that the majority of newborn stars are formed in clustered environments -- bound and unbound star clusters, and stellar associations \citep{Lada2003}. However, most of the clusters and associations eventually dissolve into the galactic stellar field, thereby contributing to the build-up of the global stellar populations of galaxies \citep{PortegiesZwart2010}. Thus, studies of star clusters are key in finding solutions to unanswered questions about star formation. Knowledge of clusters' physical (age, mass, metallicity, and extinction) and structural (size and ellipticity) parameters provides us with an insight into the formation and evolution of their host galaxy discs, and enables us to better comprehend general star formation processes. 

Star cluster studies in the Milky Way disc are rather limited due to the fact that the Solar System resides near the galactic plane. However, the proximity of other disc galaxies in our neighbourhood, such as Andromeda (\object{M\,31}) or Triangulum (\object{M\,33}), combined with the outstanding capabilities of $Hubble$ Space Telescope (HST) imaging, provide us with an excellent opportunity to study a large number of unresolved and semi-resolved star clusters located in the discs of these galaxies.

Parameters of extragalactic star clusters are derived using various methods and algorithms. Structural parameters of unresolved or partially resolved star clusters are most commonly evaluated by fitting 2D models to the observed surface brightness distributions \citep{Larsen1999, Narbutis2014, Brown2021}. Meanwhile, for star cluster parameter derivation, several families of methods are used, all of which exhibit varying accuracy and applicability limits and caveats. Many currently applied techniques for star clusters in the local Universe ($\lesssim$10\,Mpc) rely on broad-band photometry. The most accurate methods are based on constructing clusters' colour-magnitude diagrams (CMDs) from individual member stars and comparing them to the synthetic CMDs produced by simulating stellar populations or fitting theoretical isochrones \citep{Weisz2015, Johnson2016, Wainer2022, Ceponis2024}. However, such methods require that individual upper main-sequence (MS) stars are well resolved as it is necessary to obtain reliable measurements of stars below the MS turn-off (MSTO) point. Therefore, accurate and robust results are achievable only for relatively nearby and young star clusters, greatly limiting the number of objects that can be studied using CMD fitting methods; \citet{Wainer2022} have shown that parameter estimates are good up to 300\,Myr and for clusters more massive than 1000\,$\rm{M_\sun}$. Integral spectroscopy methods \citep{Caldwell2009, Caldwell2011, Caldwell2016} also provide accurate cluster parameter estimates, but given the capabilities of present-day observatories, only studies of rather massive extragalactic clusters can be performed.

Another group of popular methods employed to derive cluster parameters fit stochastic theoretical stellar population models to the observed integrated cluster magnitudes and colour indices (CIs) \citep{Deveikis2008, Fouesneau2010, Fouesneau2014, deMeulenaer2013, Krumholz2015}. These techniques are applicable to study lower-mass, older, and more distant star clusters, greatly increasing the number of objects to be investigated. This enables us to study the star formation histories of galactic discs based on a wider range of star cluster masses, and on longer timescales. However, when such methods are used, the accuracy of the derived star cluster parameters strongly depends on the uncertainties of aperture photometry, and on whether projected background and foreground stars  (hereinafter field stars), and the brightest evolved member stars are properly accounted for. Especially in the low-mass range ($\lesssim$3000\,$\rm{M_\sun}$), integrated photometry methods are extremely sensitive to stochastic effects of star clusters -- mostly the stochastic sampling of the stellar initial mass function (IMF) -- especially at young ages ($<$1\,Gyr) \citep{Fouesneau2010, Beerman2012, Anders2013, deMeulenaer2013, deMeulenaer2017}.

Most of the stochastic variation in cluster brightness and colour in the case of low-mass objects is driven by stars that have left the MS and already are in transient late-evolutionary phases. The upper MS and evolved star stages of low-mass clusters ($\lesssim$$10^{4}\,\rm{M_\sun}$) are sparsely populated, causing cluster photometric parameters to vary in broad ranges. In many cases, these large differences are caused by only a single or a few evolved stars. \citet{Beerman2012} studied the effects of bright post-main-sequence (PMS) stars on the accuracy of star cluster parameters derived using the integrated photometry method. They showed that these evolved stars severely impact cluster measurements and the accuracy of derived ages and masses; parameters can be estimated more precisely when bright resolved PMS members are excluded from aperture photometry and only the unresolved light component is used. Furthermore, \citet{deMeulenaer2017} notice that the accuracy of cluster physical parameter derivation is greatly affected by bright field stars. They show that in cases when bright field stars (which are mostly red) reside within the aperture used for the photometry of the star cluster, strong age-extinction degeneracies are introduced. 

To tackle the aforementioned problems arising due to the bright field stars projecting within the applied aperture and the stochasticity of the brightest evolved star cluster members, \citet{Naujalis2021} have presented a new adaptive aperture photometry method. The basis of this novel approach is that measurements of CIs are performed using smaller apertures, containing only the central part of the cluster, reducing the number of field contaminants by a few times. This technique was applied by \citet{Naujalis2021} to 1181 star clusters from the \object{M\,31} Panchromatic $Hubble$ Andromeda Treasury (PHAT; \citealt{Dalcanton2012}) survey cluster catalogue \citep{Johnson2015}. Precisely following the procedures described by \citet{Naujalis2021}, a study by \citet{Krisciunas2023} provides photometry results of the additional 1477 PHAT clusters to supplement the sample of star clusters presented by \citet{Naujalis2021}. Both studies show that when compared to the conventional aperture photometry approach, the proposed new method gives more robust results in the sense of more consistent CIs as well as more compatible photometry with the stochastic star cluster models.

Although integrated cluster photometry appears to be an excellent tool for studying large samples of star clusters in the local Universe, deriving their parameters and investigating star formation histories of disc galaxies, the accuracy of the method and its limitations due to stochastic and environmental effects still remain to be understood. As we seek to comprehend the formation and evolution of galaxies, it is necessary to understand the intrinsic precision limits of such methods, so that we do not over-interpret data.

In this study, we aim to estimate achievable accuracy limits of the aperture photometry methods for star cluster studies in the local Universe. We present an algorithm for modelling stochastic 3D star clusters and their HST images. It was employed to generate a large grid of cluster models that covers the parameter space of real clusters observed in \object{M\,31}. The simulated 3D clusters were then projected onto a 2D plane from differing viewing angles, to imitate their observations from 100 directions. The images were modelled in six passbands to be consistent with real observations of \object{M\,31} star clusters in the PHAT survey. Simulated frames were then measured using aperture photometry, as has been described by \citet{Naujalis2021} and \citet{Krisciunas2023}. Furthermore, the results of the aperture photometry were then used to run cluster parameter determination tests by applying the algorithm described in \citet{deMeulenaer2017}. Also, the fact that we had simulated images of the same clusters observed from various angles allowed us for the first time to determine uncertainties in the cluster aperture photometry introduced due to the randomness of an object's orientation towards our line of sight.

We emphasise that all experiments in this study were performed under idealised conditions -- there is no sky background in the star cluster images. This means that all uncertainties and distorting effects discussed in this paper are at their minimum. The inclusion of any precision-limiting factors, such as the sky background or bright resolved field stars, would decrease reported accuracies, especially in the case of low-mass clusters. In other words, here we present the most optimistic accuracy limits of aperture photometry for studying star clusters in the local Universe. Nonetheless, it is important to know the upper limits of the derived cluster parameter precision using aperture photometry methods; it can help one to better understand the accuracy limits when discussing star formation bursts or other evolutionary processes in nearby galaxy stellar populations.

The structure of this paper is as follows: in Sect.\,\ref{Sec2} we present an algorithm for simulating 3D star clusters and modelling their images; in Sect.\,\ref{Sec3} we briefly describe procedures of adaptive aperture photometry; in Sect.\,\ref{Sec4} we present and discuss results; and finally in Sect.\,\ref{Sec5} we summarise our study.
 
\section{Artificial star clusters}
   \label{Sec2}

\begin{figure}
        \centering
        \includegraphics[width=9cm]{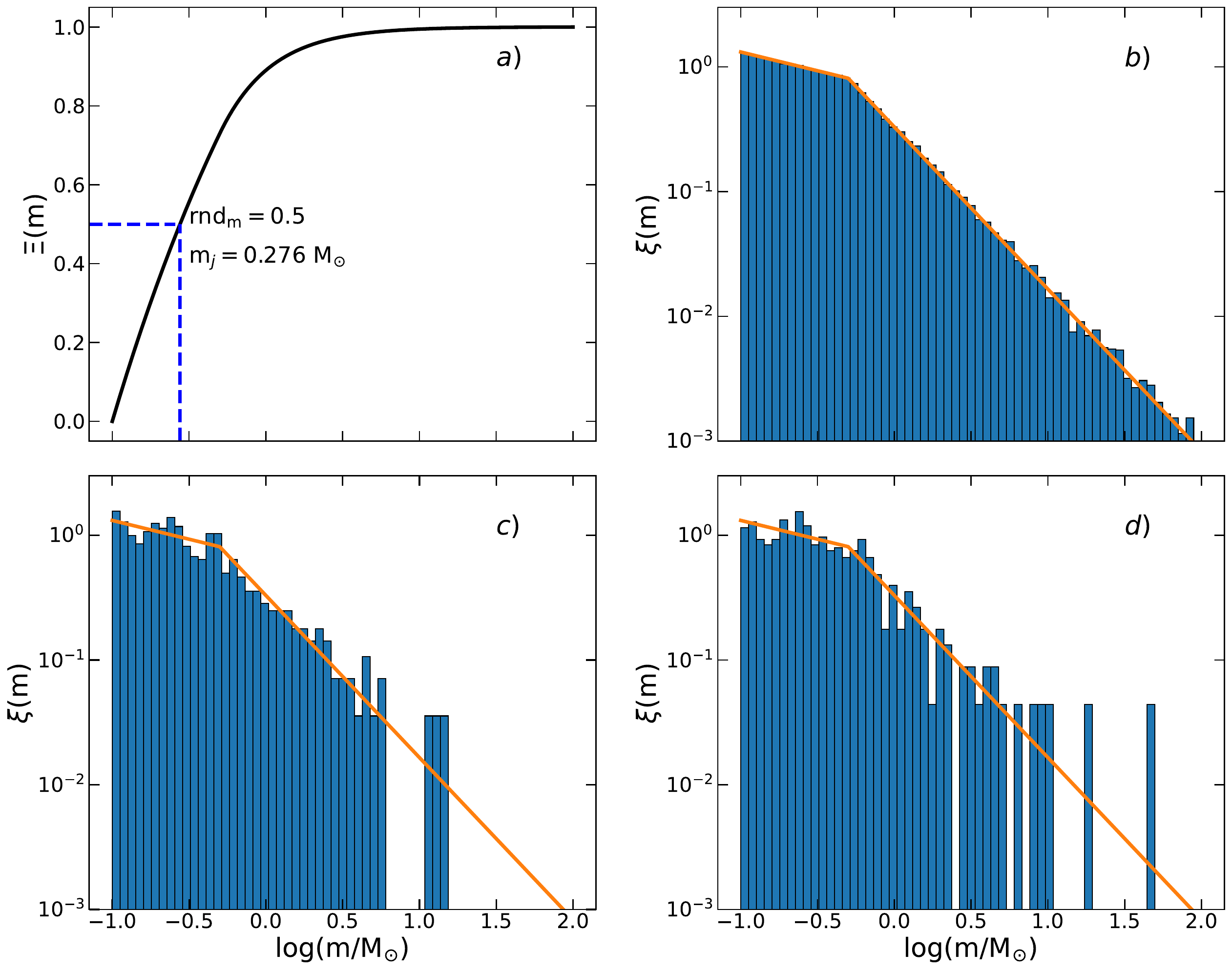}
        \caption{Stochastic stellar mass sampling of individual cluster stars. a) Cumulative mass distribution $\Xi(\rm m)$ in the range from 0.1 to 100\,$\rm M_{\sun}$ (black line) and its application for stellar mass sampling (blue dashed line). b) Histogram -- log$\rm(M/M_{\sun})=5.0$ cluster mass function matches a theoretical IMF with high accuracy -- weak stochastic effects. c) and d) Histograms -- two different generations of log$\rm(M/M_{\sun})=2.5$ cluster -- low-mass clusters exhibit significant mass function stochasticity in the tail of massive stars. b-d) The orange lines represent the IMF $\xi(\rm m)$ for an infinite mass cluster (probability density function).}
        \label{fig1}
\end{figure}
\begin{figure}
        \centering
        \includegraphics[width=9cm]{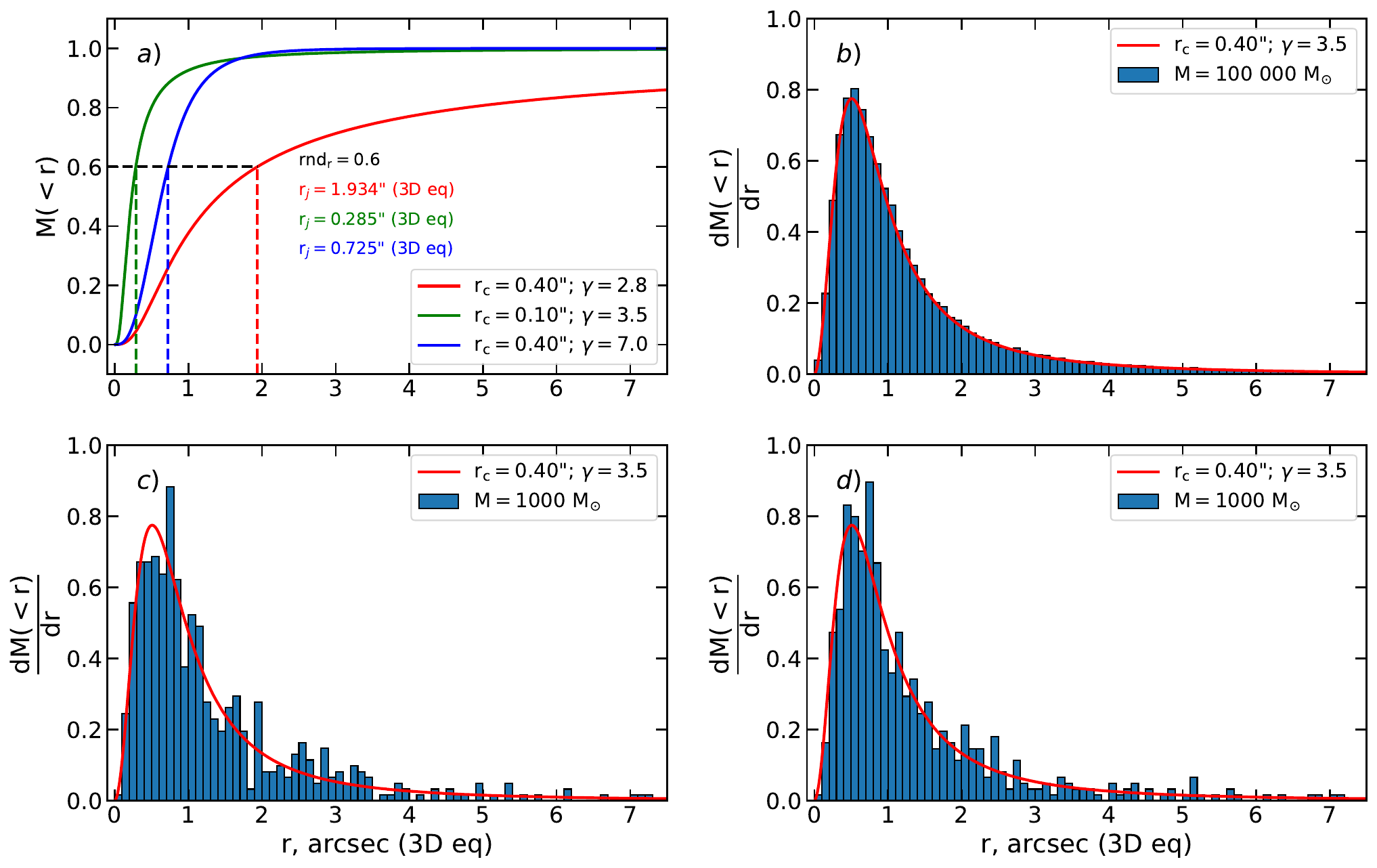}
        \caption{Sampling 3D distances of individual stars from the cluster centre. a) Three cumulative radial cluster mass profiles with different EFF profile parameters; distances sampled with $\rm rnd_{r}=0.6$ are shown for each profile. b) 3D distance sampling result for log$\rm(M/M_{\sun})=5.0$ cluster. c) and d) Two different log$\rm(M/M_{\sun})=3.0$ cluster generations. In the case of low-mass clusters, the stellar spatial distribution exhibits significant stochastic effects. Red lines in b)-d) mark a distance sampling probability density function ${d{\rm M(<r)}}/{d{\rm r}}$.}
        \label{fig2}
\end{figure}

We developed an algorithm for simulating artificial 3D star clusters and their images when projecting these clusters onto a 2D plane from various different angles. It is based on $\tt SimClust$ code \citep{Deveikis2008} developed for 2D star cluster simulation. All parts of the mock cluster modelling were implemented using scripts written in $\tt Python$. In this section we describe the main procedures of our algorithm.

Firstly, the initial masses of individual stars are generated fully stochastically. \citet{Grudic2023}, based on their radiation magnetohydrodynamics simulations of low-mass star clusters, show that even though star formation is not random and uncorrelated, stochastic sampling appears to lead to a reasonable description of the end result of star formation; thus our choice of using fully stochastic sampling in this study is justified. Masses are stochastically sampled according to the IMF, which defines the number of stars born within a certain mass range. In this study, we used the  \citet{Kroupa2001} IMF, defined by the multi-power law:
\begin{ceqn}
\begin{align}
        \xi(m)=k \cdot b_{i} \cdot {\rm m}^{-\alpha_{i}},
\end{align}
\end{ceqn}
where $\alpha_{i}$ -- the IMF slope for the mass range of $[{\rm m}_{i-1}, {\rm m}_{i}]$, $k$ and $b_{i}$ -- normalisation and function continuation constants, respectively.

Initial stellar masses are generated in the interval from 0.1 to 100 $\rm{M_\sun}$. In this mass range, the Kroupa IMF becomes a simple equation, defined by two power-law slopes:

\begin{ceqn}
\begin{equation}
\begin{cases}
\alpha_{\text{1}} = 1.3,\quad \rm m < 0.5~\rm{M_\sun}; \\
\alpha_{\text{2}} = 2.3,\quad \rm m \geq 0.5~\rm{M_\sun}.
\end{cases}
\end{equation}
\end{ceqn}
In order to sample individual stellar masses, a cumulative mass distribution function, which gives a probability that a certain mass, $\rm m' \leq m$, is constructed by integrating Eq. 1:

\begin{ceqn} 
\begin{align}
        \label{eq:3} \Xi({\rm m})= \int_{{\rm m}_{min}}^{\rm m} \xi({\rm m'})\,{\rm dm'},
\end{align}
\end{ceqn}
$k$ and $b_{i}$ constant values are selected such that Eq.\,\ref{eq:3} is continuous and its value varies from 0 to 1 in a given stellar mass interval. The mass of a single star is sampled by generating a random number ${\rm rnd_m}$ from the uniform distribution in the range from 0 to 1, equating this value to $\Xi({\rm m})={\rm rnd_m}$, and then applying the inverse function:
\begin{ceqn}
\begin{align}
        {\rm m}_{j}=F^{-1}(\Xi({\rm m})={\rm rnd_m}),
\end{align}
\end{ceqn}
where ${\rm m}_{j}$ is the initial stellar mass of the $j$-th star in the cluster.
A graphical illustration of the described method for generating stellar masses is presented in Fig.\,\ref{fig1}.

Positions of stars inside the cluster are distributed according to an empirical Elson-Fall-Freeman \citep[EFF;][]{Elson1987} profile. The use of the EFF model is motivated by our desire to mimic real young clusters as close as possible, because such objects are the most susceptible to uncertainties arising due to bright field stars and stochastic effects. This choice is well-supported by the fact that a number of studies \citep[e.g.][]{Elson1987, Larsen1999, Larsen2004, McLaughlin2005, Bastian2012, Cuevas2020, Brown2021} show that EFF-type profiles describe observed radial profiles of young star clusters accurately. In the case of the EFF model, the radial star density profile is described by the following function:
\begin{ceqn}
\begin{align}
        \label{eq:5} \rho(\rm r)=\rho_{0}\left( 1+\dfrac{\rm r^{2}}{\rm r_{0}^{2}}\right)^{-(\gamma+1)/2},
\end{align}
\end{ceqn}
where $\rho_{0}$ -- central cluster density; $\rm r$ -- de-projected (3D) distance from the cluster's centre; $\rm r_{0}$ -- scale factor; $\gamma$ -- power-law slope, which determines the behaviour of a radial profile in the cluster's outer regions.

Stellar spatial distribution is generated using the method suggested by \citet{Aarseth1974} for modelling initial coordinates of stars in dynamical N-body cluster simulations. It is noteworthy that this approach assumes that all cluster member stars are of the same mass. This method requires a cumulative radial cluster mass profile, which can be derived by integrating Eq.\,\ref{eq:5}:
\begin{ceqn}
\begin{equation}
        \begin{aligned}[b]
                &\rm M(\rm <r)=\int_{0}^{\rm r} 4\pi x^2 \rho(x)\,dx= \\ 
                &\frac{4\pi\rho_0}{3}  \rm r^3 \cdot {}_2{\textit F_1}\left(\frac{3}{2};\frac{\gamma+1}{2};\frac{5}{2};-\frac{\rm r^2}{\rm r_0^2}\right).
        \end{aligned}
\end{equation}
\end{ceqn}
where ${}_2F_1(a;b;c;z)$ is the Gauss hypergeometric function. The function $\rm M(<r)$ is normalised so that its values vary from 0 to 1 and then a random number ${\rm rnd_r}$ from a uniform distribution is generated in this range. Just as in the stellar mass sampling scheme, ${\rm rnd_r}$ are attributed to ${\rm M(<r)}$ = $\rm rnd_{r}$, and then an inverse function is applied to derive 3D distances of individual stars from the cluster's centre, ${\rm r}_{j}$:
\begin{ceqn}
\begin{align}
        {\rm r}_{j}=F^{-1}({\rm M(<r)}={\rm rnd_r}).
\end{align}
\end{ceqn}
The method for generating 3D stellar distances is presented graphically in Fig.\,\ref{fig2}.

The next step of the algorithm is to assign coordinates to stars inside the cluster. To do that, we generated two random numbers from uniform distributions -- $u_{j}$ (in the ranges from 0 to 2$\pi$) and $v_{j}$ (from $-1$ to 1). These numbers then were transformed into spherical coordinates of stars by assuming ${\rm r}_j = 1$:
\begin{ceqn}
        \begin{equation}
                \begin{cases}
                        \phi_{j}=u_{j}; \\
                        \theta_{j} =\cos^{-1}v_{j}.
                \end{cases}
        \end{equation}
\end{ceqn}
The spherical coordinates were then converted to Cartesian coordinates:
\begin{ceqn}
        \begin{equation}
                \begin{cases}
                        {\rm X}_{j}={\rm r}_{j}\cdot\sin\theta_{j}\cdot\cos\phi_{j}; \\
                        {\rm Y}_{j}={\rm r}_{j}\cdot\sin\theta_{j}\cdot\sin\phi_{j}; \\
                        {\rm Z}_{j}={\rm r}_{j}\cdot\cos\theta_{j}.
                \end{cases}
        \end{equation}
\end{ceqn}

The algorithm for sampling individual stars (their initial masses and spatial distributions) was repeated until a cluster's target mass was reached or exceeded inside a sphere of a 7.5\,arcsec radius (at the  distance of \object{M\,31} it is $\sim$28.5\,pc). When the algorithm halts, it is evaluated whether total generated mass inside the sphere is closer to the targeted cluster mass with the inclusion of the last sampled star, or excluding it. The cluster that deviates less from the planned mass is taken as the final result of the modelling.

Once stellar masses and spatial distributions are generated, magnitudes need to be assigned according to star masses and the age of a cluster. To do so, we interpolated the PARSEC-COLIBRI \citep{Marigo2017} isochrones\footnote{\url{http://stev.oapd.inaf.it/cgi-bin/cmd}}. In this study, we assumed a fixed solar metallicity and zero interstellar extinction.

Since we aim to simulate accurate images of artificial clusters that match images of \object{M\,31} star clusters observed in the PHAT survey \citep{Dalcanton2012}, isochrones were interpolated in six corresponding HST passbands, covering a spectral range from 0.25 to 1.6\,$\mu$m. Before interpolation, absolute magnitudes given in isochrones were converted to apparent ones by applying the \object{M\,31} distance modulus of $(m-M)_{0}=24.47$ \citep{McConnachie2005}.

\begin{table}
        \setlength{\tabcolsep}{4pt}
        \caption{\label{t2} Artificial star cluster parameters.}
        \centering
        \begin{tabular}{ll}
                \hline\hline
                Parameter & Values\\
                \hline
                log$\rm(M/M_{\sun})$ & 2.0, 2.5, 3.0, 3.5, 4.0\\
                log$\rm (t/yr)$ & 7.0, 8.0, 9.0, 10.0\\
                $\rm r_{c}$, arcsec\tablefootmark{a} & 0.05, 0.10, 0.20, 0.40, 0.80\\
                $\gamma$ & 2.2, 2.4, 2.8, 3.5, 7.0\\
                \hline
        \end{tabular}
        \tablefoot{\tablefoottext{a}{Cluster core radius, as defined by \citet{King1962}; the relation of the core radius to the EFF scale factor: ${\rm r_{c}}={\rm r_{0}}(2^{\sfrac{2}{\gamma}}-1)^{\sfrac{1}{2}}$}.
        }
        \label{table:tab2}
\end{table}
\begin{figure}
        \centering
        \includegraphics[width=9.0cm]{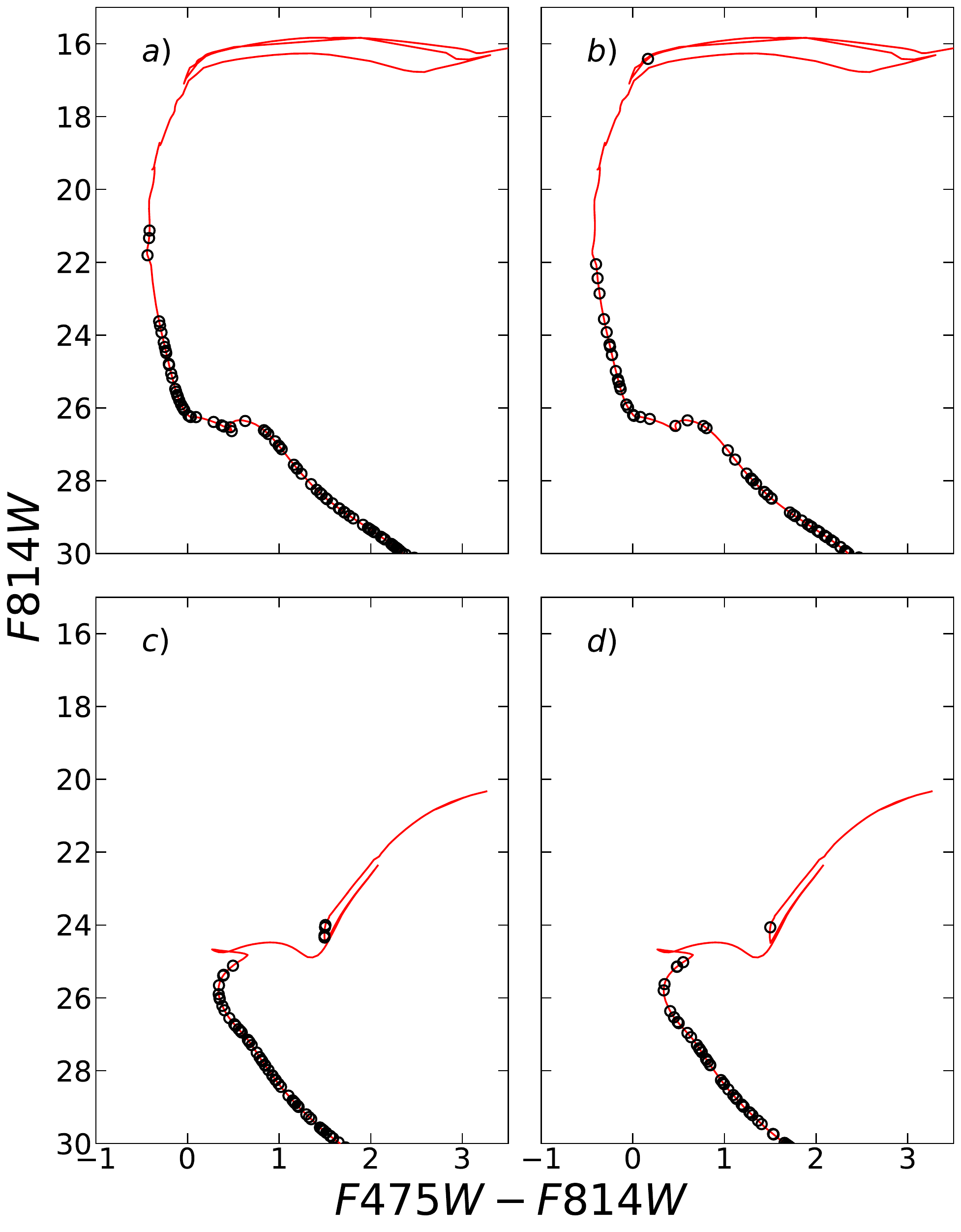}
        \caption{CMDs of two simulated clusters of mass log$\rm(M/M_{\sun})=2.5$ and of age log$\rm (t/yr)=7.0$ (upper row) and log$\rm (t/yr)=9.0$ (bottom row). The generation in the left column is the same as shown in Fig.\,\ref{fig1}c, meanwhile the  right column matches Fig.\,\ref{fig1}d. Red lines mark PARSEC-COLIBRI isochrones. Open circles -- individual stars.}
        \label{fig3}
\end{figure}

The CMDs in Fig.\,\ref{fig3} show two artificial clusters of the same mass at two different ages together with appropriate isochrones. This illustrates that in the case of young low-mass clusters, even though the physical parameters are the same, clusters exhibit significant differences due to the effects of IMF stochasticity.


We generated a grid of artificial 3D star cluster models, covering the parameter space of real clusters observed in \object{M\,31}, 500 grid nodes in total (see Table 1). In each node we simulated 100 independent clusters and each cluster projected with 100 different viewing angles. Therefore, taking into account six PHAT passbands, we ended up with $3\cdot 10^7$ images. 

Realistic images of artificial clusters were modelled using a similar approach as in \citet{Bialopetravicius2019}. To model images of individual stars we made use of $\tt TinyTim$\footnote{\url{https://www.stsci.edu/hst/instrumentation/focus-and-pointing/focus/tiny-tim-hst-psf-modeling}} package, which enables the modelling of point spread functions (PSFs) for different HST instrument and passband combinations. An individual star was modelled by multiplying its total flux in any of the passbands by the respective $\tt TinyTim$ PSF. This gave pixel flux values in ${\rm e^-}$/s. Total fluxes of stars in any photometric passband ('?') were calculated according to the following equation:
\begin{ceqn}
        \begin{align}
                {\rm FLUX}_{F?W}=10^{-0.4\cdot (F?W-F?W_0)}, 
        \end{align}
\end{ceqn}
where $F?W$ is a magnitude in any of the PHAT passbands and $F?W_0$ is a corresponding zeropoint. Appropriate zeropoints for the ACS camera were calculated using the ACS zeropoints calculator\footnote{\url{https://www.stsci.edu/hst/instrumentation/acs/data-analysis/zeropoints}}; zeropoints for both channels of WFC3 were taken from the Space Telescope Science Institute (STScI)\footnote{\url{https://www.stsci.edu/hst/instrumentation/wfc3/data-analysis/photometric-calibration}}. Images of artificial clusters were simulated within frames of $15\times15$\,arcsec ($\sim$$57\times57$\,pc at the distance of \object{M\,31}). Images of each simulated cluster were produced using 100 different 2D projections, imitating  observations from 100 angles. The same 100 random projection angles were used for all artificial clusters. The modelled cluster images were saved in the FITS file format.

\section{Adaptive aperture photometry}
        \label{Sec3}

\citet{Naujalis2021} introduced a novel adaptive aperture photometry method of mitigating problems of star cluster parameter derivation arising due to the effects of bright field stars and evolved bright cluster members, which were noticed by \citet{deMeulenaer2017} and \citet{Beerman2012}, respectively. This method was applied to the PHAT star cluster catalogue \citep{Johnson2015} objects by \citet{Naujalis2021} and \citet{Krisciunas2023}. Both of these studies show that adaptive aperture photometry techniques give more consistent CIs of star clusters when compared to the conventional approach, which makes them promising for studies of semi-resolved and unresolved clusters in the local Universe.

Since one of the goals of this study is to determine whether assumptions of adaptive aperture photometry are correct, in this section, we briefly describe the aforementioned method (for a more extensive explanation, refer to the publications by \citet{Naujalis2021} and \citet{Krisciunas2023}).

The key idea of adaptive aperture photometry is measuring a star cluster with two co-centred apertures. The larger one called T ('total') is meant to measure the cluster's total flux. The smaller one called C ('colour') is selected in such a manner that it encircles the central part of the cluster and avoids objects that can lead to incorrectly derived CIs by falling within the T aperture -- bright field and/or evolved cluster members. Measurements in the C aperture provide less contaminated and more colour-consistent cluster fluxes. It is noteworthy that, in the studies by \citet{Naujalis2021} and \citet{Krisciunas2023}, in the majority of cases a rule is followed, that C aperture radii ($\rm {R}_{C}$) have to be equal to or larger than the half-light radii of corresponding clusters.

As a base to derive final colour-consistent cluster magnitudes, \citet{Naujalis2021} chose the $F475W$ passband, since in the PHAT survey, observations in this passband are of the highest signal-to-noise ratio and are less contaminated by the light of resolved field stars from old \object{M\,31} populations, when compared to the other passbands. In this study, we follow the procedures of the adaptive aperture photometry precisely as presented by \citet{Naujalis2021}.

Final colour-consistent cluster magnitudes in all passbands are derived by applying aperture correction (AC), determined for the cluster in the $F475W$ passband, $F475W_{\rm AC}$. It is defined as the difference of $F475W$ magnitudes when measured through T and C apertures: ${F475W}_{\rm AC} = {F475W}_{\rm T} - {F475W}_{\rm CA}$. The final magnitudes (C) are ${F?W}_{\rm C} = {F?W}_{\rm CA} + {F475W}_{\rm AC}$, where the '?' marks the HST three-digit passband code and ${F?W}_{\rm CA}$ are the magnitudes measured through the C aperture.

As \citet{Naujalis2021} and \citet{Krisciunas2023} notice, it is important to stress that such magnitude transformation is reliable only for clusters without strong colour gradients beyond $\rm {R}_{C}$. These can be introduced due to a number of reasons; for example, mass segregation of stars or real gradients of stellar populations within some clusters.

\section{Results and discussion}
        \label{Sec4}
        
\begin{figure}
        \centering
        \includegraphics[width=9cm]{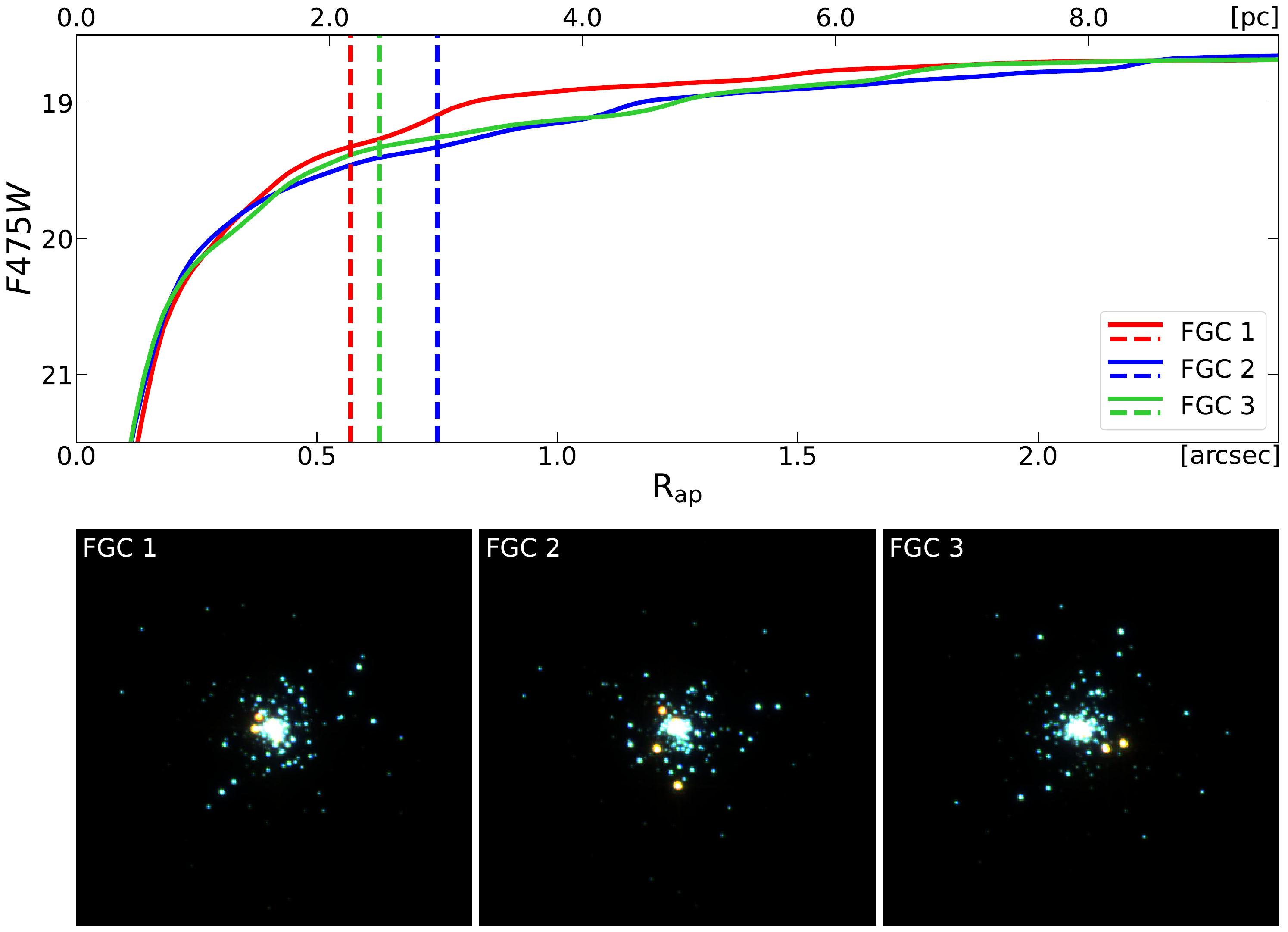}
        \caption{FGCs in the $F475W$ passband for the artificial star cluster (log${\rm (t/yr)}=8.0$, log$\rm(M/M_{\sun})=3.5$, $\rm r_{c}=0.2$\,arcsec, and $\gamma=2.8$) observed from three viewing directions. The colour images (15$\times$15\,arcsec) were produced by combining $F336W$, $F475W$, and $F814W$ data together with the help of the $\tt APLpy$ package \citep{aplpy2012} and assuming the distance of \object{M\,31}. Dashed vertical lines mark half-light radii measured for respective projections.}
        \label{fig4}
\end{figure}

To investigate stochastic effects on star cluster aperture photometry results and to determine the  applicability limits of this method, we measured artificial cluster images following the recipes used by \citet{Naujalis2021} and \citet{Krisciunas2023}. All procedures of aperture photometry were completed using tools provided in the $\tt Python$ package $\tt photutils$\footnote{\url{https://doi.org/10.5281/zenodo.596036}}.

Radial flux-growth-curves (FGCs) were measured up to the aperture with a radius of 7.0\,arcsec in all modelled passbands, using an aperture radius step of 0.02\,arcsec. The FGCs enabled us to derive fluxes and magnitudes, as well as CIs, of artificial star clusters when measured through apertures of various sizes. Also, FGCs allow us to determine the radii of apertures that enclose various percentages of the clusters' full flux, ${\rm R}_{\rm X}$ (also referred to as cluster structural parameters' R further on in this paper), where X is the percentage of flux enclosed inside the circular aperture of radius ${\rm R}_{\rm X}$. 

In Fig.\,\ref{fig4}, we show FGCs and colour images of the same simulated cluster when observed from three different directions. It is evident that the direction of observations has a significant influence on the shape of FGCs and measured structural parameters. The analysis of FGCs enabled us to investigate effects of random cluster orientation on aperture photometry and structural parameters. 

Further on in this paper, when we present various effects depending on cluster mass and age, we mostly show results for clusters with fixed EFF profile parameters ${\rm r}_{\rm c}=0.20$\,arcsec and $\gamma=2.8$. This is motivated by the fact that this particular EFF profile is consistent with typical parameters reported for real star clusters in nearby galaxies \citep{Larsen1999, Scheepmaker2007, Sableviciute2007, Mora2009, Vansevicius2009, Bastian2012, Brown2021}. Also, when the effects depending of ${\rm r}_{\rm c}$ and $\gamma$ are shown, we mostly take results for 100\,Myr old clusters with an initial mass of 1000\,$\rm{M_\sun}$. These are the typical cluster age and mass values reported for star clusters in the \object{M\,31} disc \citep{Vansevicius2009, Fouesneau2014, deMeulenaer2017}.

\subsection{Photometric cluster parameters}
\label{Sec41}

\begin{figure}
        \centering
        \includegraphics[width=9cm]{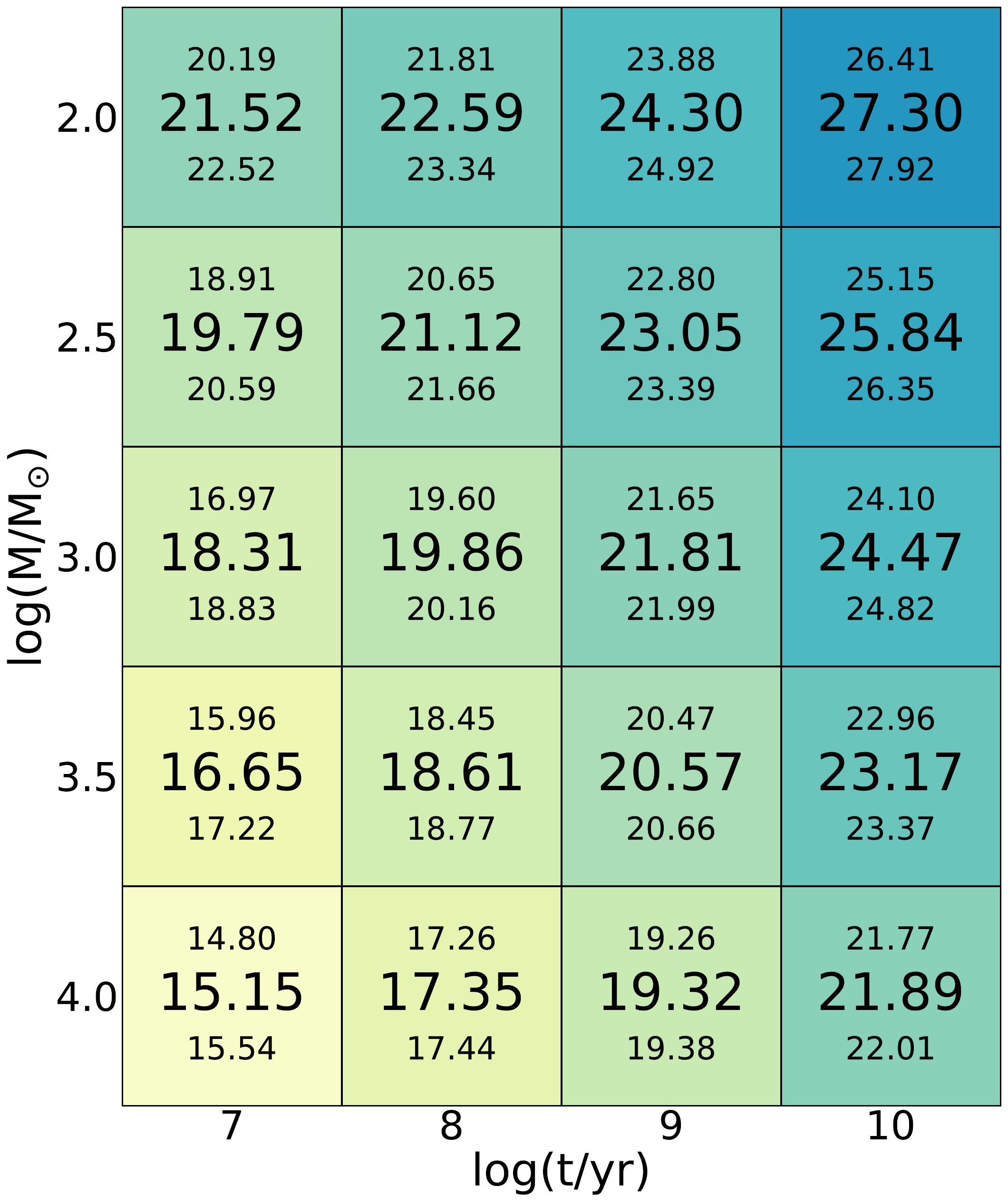}
        \caption{Apparent $F475W$ magnitudes of star clusters at the distance of \object{M\,31} depending on their mass and age. The central value in each cell represents the median $F475W$ magnitude for a given mass and age combination; meanwhile, the upper and lower values show the 16th and 84th percentiles, respectively. Magnitudes are measured using a circular aperture of radius $\rm R_{ap}=7.0$\,arcsec.} 
        \label{fig5}
\end{figure}
\begin{figure}
        \centering
        \includegraphics[width=9cm]{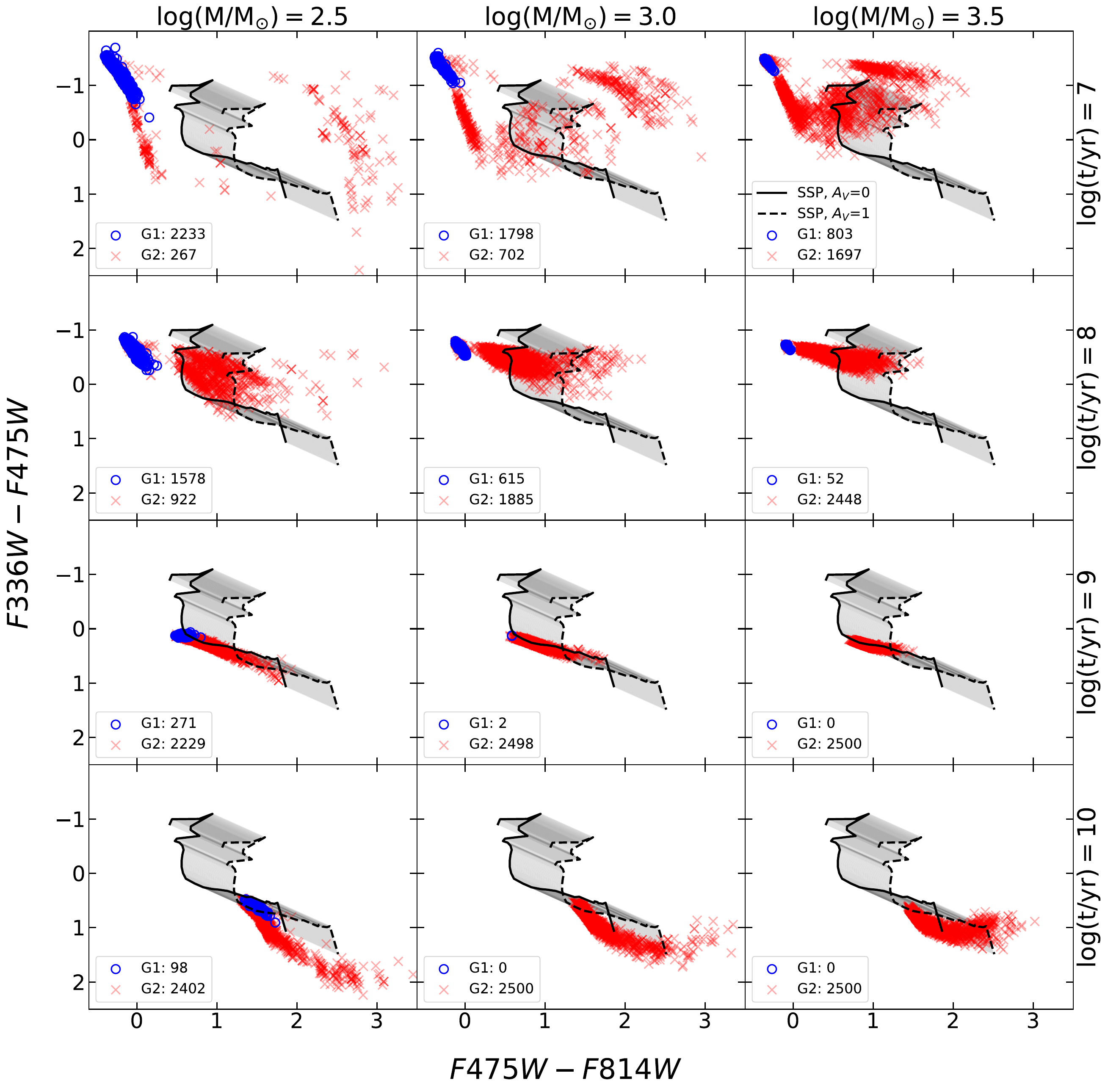}
        \caption{TCDs of simulated stochastic star clusters. Each panel shows aperture photometry results for different combinations of cluster mass (marked above the panels) and age (marked to the right of the panels). For the definitions of the G1 and G2 cluster groups, see the text. Black lines (solid -- no interstellar extinction; dashed -- $A_{V}=1$, calculated using the \citet{Fitzpatrick1999} extinction curve) -- PARSEC-COLIBRI based non-stochastic SSP solar metallicity models \citep{Marigo2017} from 10\,Myr to 12.6\,Gyr. Measurements are performed using the circular aperture of radius $\rm R_{ap}=3.0$\,arcsec.} 
        \label{fig6}
\end{figure}
\begin{figure}
        \centering
        \includegraphics[width=9cm]{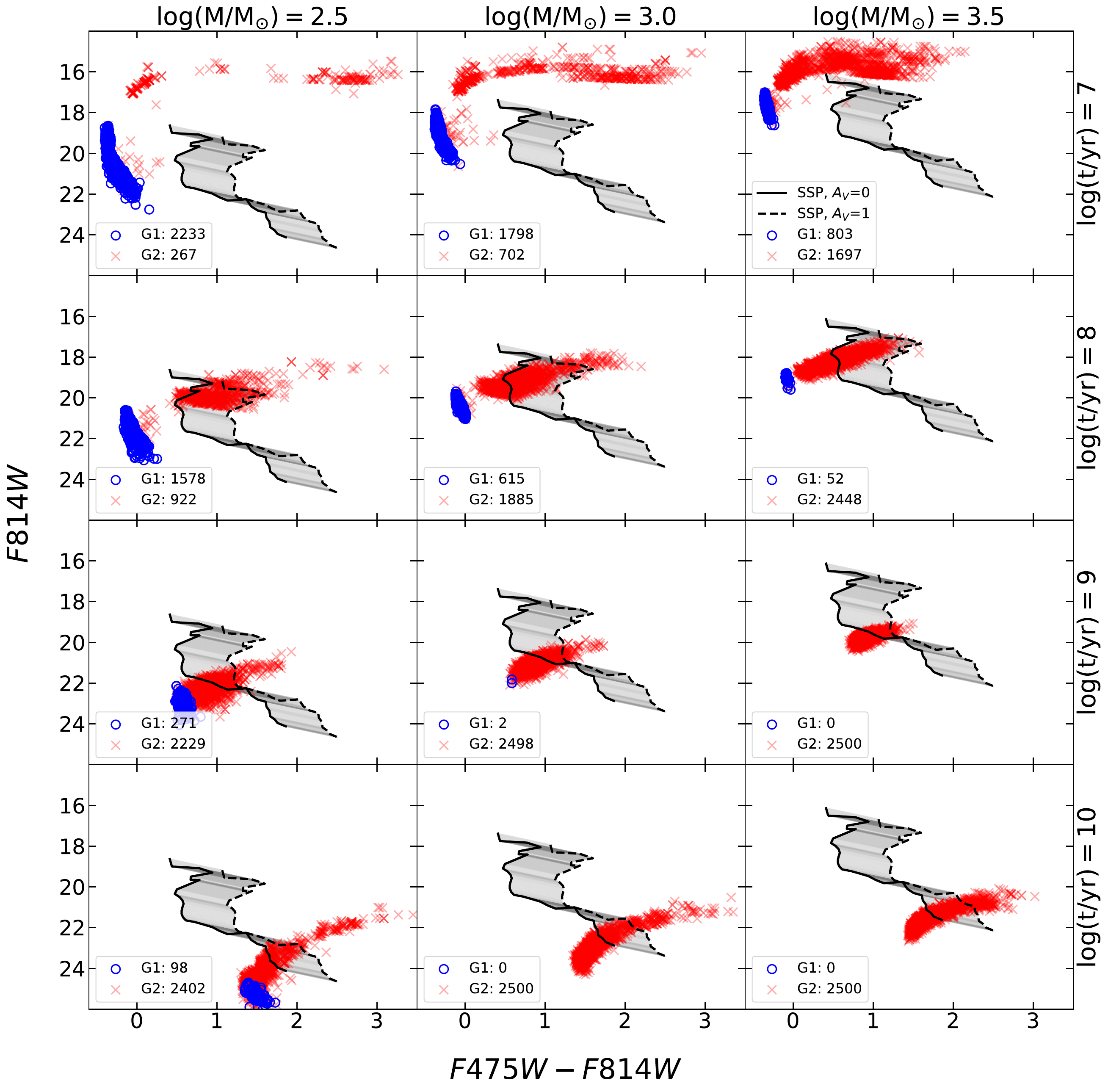}
        \caption{Same as in Fig.\,\ref{fig6}, but CMDs are shown.} 
        \label{fig7}
\end{figure}
\begin{figure}
        \centering
        \includegraphics[width=9cm]{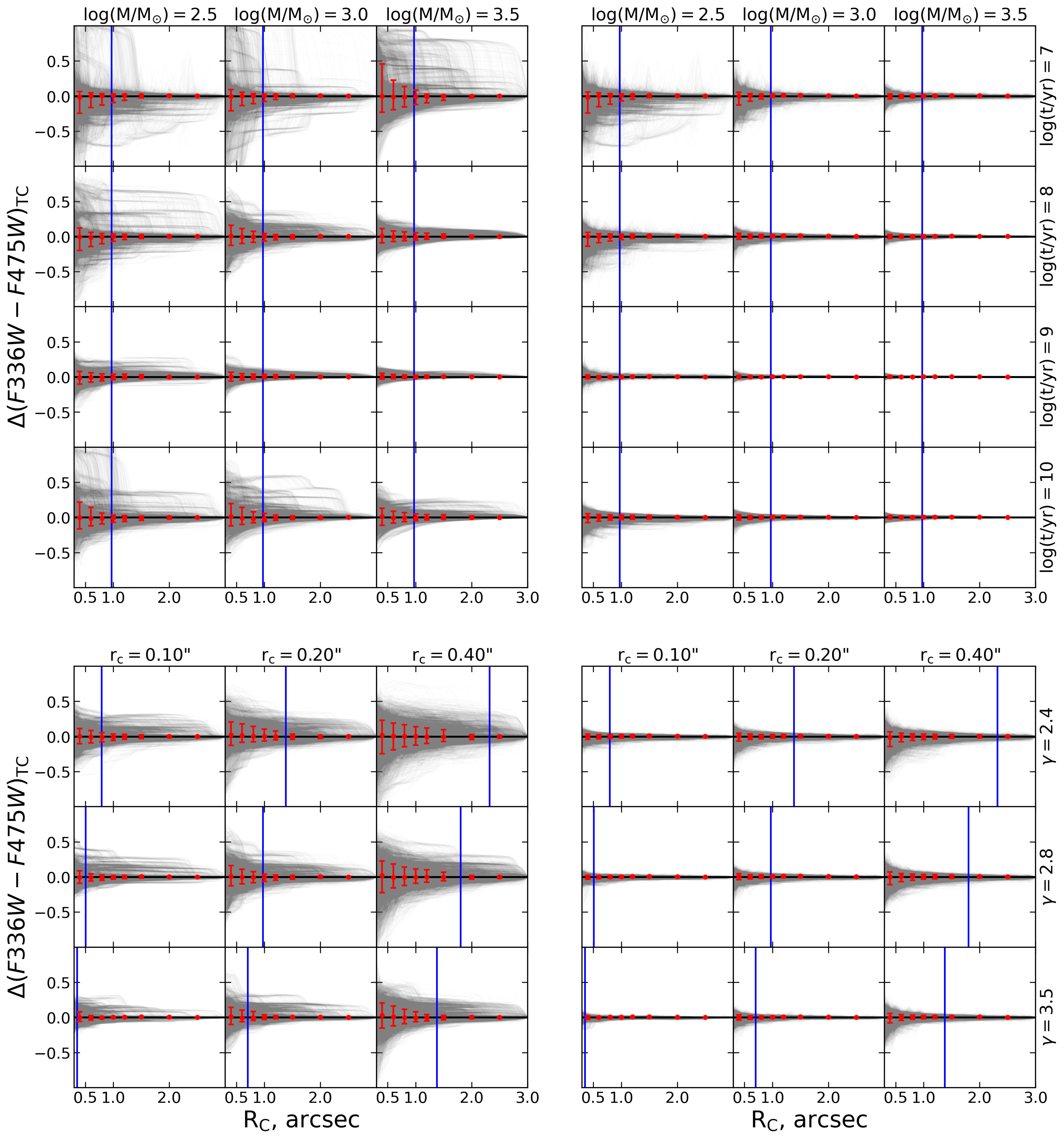}
        \caption{Differences of $F336W-F475W$, when measured through T and C apertures. The differences are shown as grey lines. The X-axis -- the radius of C aperture. The T aperture radius of $\rm R_{T}=3.0$\,arcsec is used. Red markers show the median difference for a given $\rm R_{C}$, error bars mark the 16th and 84th percentiles. The blue vertical lines mark radial distances equal to double half-light radii, 2$\cdot {\rm R_{50}}$. The top-row panels show differences depending on cluster mass and age (the cluster radial profile parameters are $\rm r_{c}=0.2$\,arcsec and $\gamma=2.8$); meanwhile, the bottom panels show differences depending on radial profile parameters (the clusters of log$\rm(M/M_{\sun})=3.0$ and log${\rm (t/yr)}=8.0$). Panel blocks on the left show results when all stars are measured, while the panel blocks on the right show results when PMS stars are subtracted from the aperture photometry results.} 
        \label{fig8}
\end{figure}
\begin{figure}
        \centering
        \includegraphics[width=9cm]{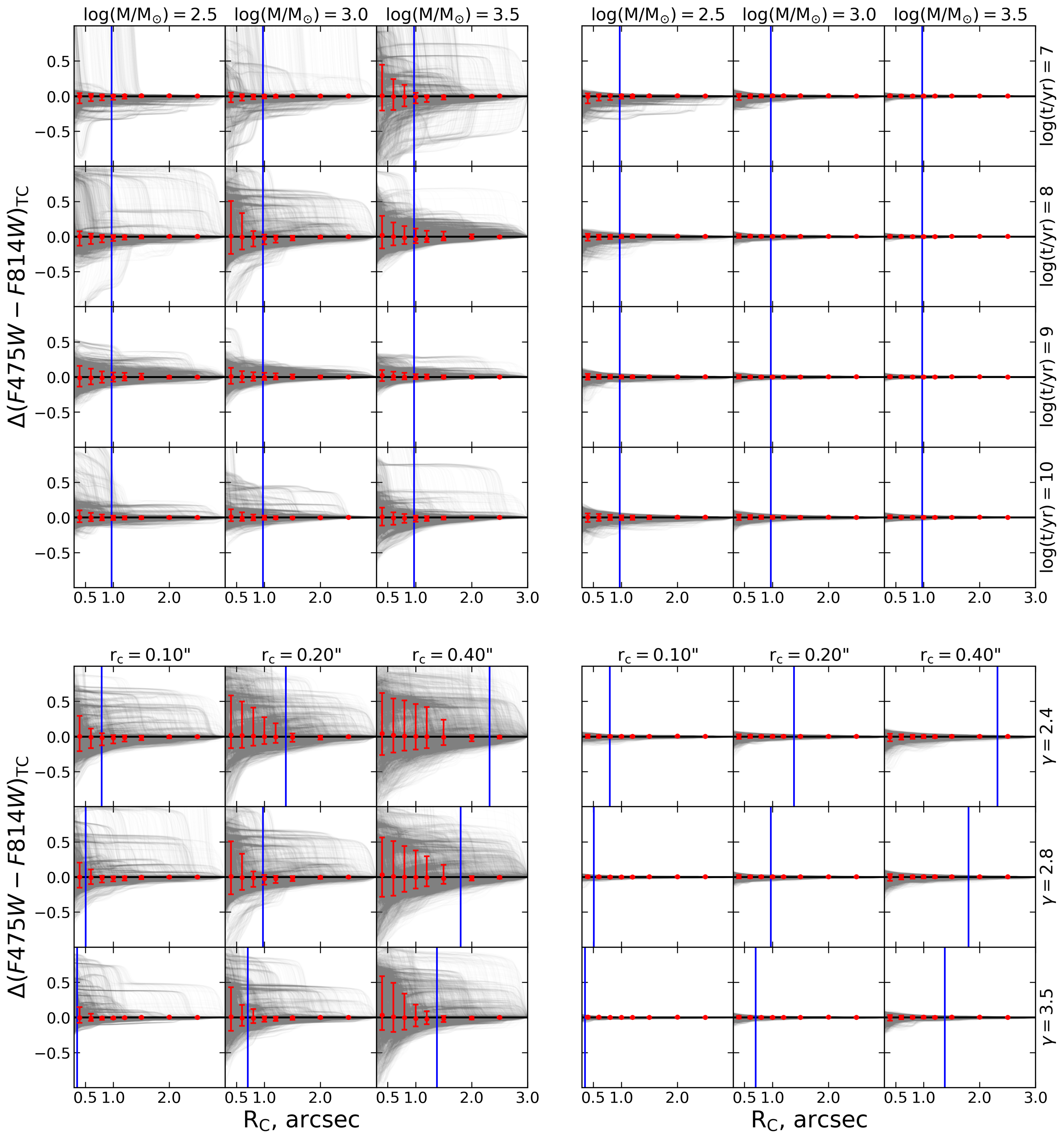}
        \caption{Same as in Fig.\,\ref{fig8}, but results for $F475W-F814W$ are shown.} 
        \label{fig9}
\end{figure}

In this section, we investigate how the stochasticity of low-mass star clusters, as well as aperture size, affects measurements of photometric parameters. The IMF stochasticity, along with the sky background determination issues, are key factors in limiting the accuracy and reliability of cluster physical parameters derived using aperture photometry. 

First of all, in Fig.\,\ref{fig5}, we show the median $F475W$ magnitude, along with the 16th and 84th percentiles, depending on cluster mass and age. At the youngest age of log$\rm (t/yr)=7.0$, clusters of log$\rm(M/M_{\sun}) \lesssim 3.5$ show a large scatter of magnitudes, as the 16-84 percentile range reaches more than 1\,mag. This hints that at such young ages mass estimates based on aperture photometry could be rather uncertain. It is noteworthy that at log$\rm (t/yr)=7.0$ magnitude ranges of the 16-84 percentiles for clusters with masses log$\rm(M/M_{\sun})=2.0$ and log$\rm(M/M_{\sun})=2.5$ slightly overlap. This suggests that mass estimation can be susceptible to large inaccuracies in this mass interval. These two cluster masses at ages of $\leq$100\,Myr are particularly interesting. In our sample, only $\sim$5\% and $\sim$10\% of clusters at the age of log$\rm (t/yr)=7.0$ with log$\rm(M/M_{\sun})=2.0$ and 2.5, respectively, have PMS stars, while at log$\rm (t/yr)=8.0$ it changes to $\sim$11\% and $\sim$37\%. This shows that in the mass and age combinations considered, such large magnitude variations (the 16-84 percentile range being from $\sim$1.0 to $\sim$2.3\,mag) are driven mostly by the stochasticity of the IMF. At higher masses and older ages, MSs are well-sampled, and sparsely populated PMS evolutionary stages become the main reason of stochastic variations. Generally, these variations tend to decrease with age up to 1\,Gyr for all studied cluster masses, and they increase once again at 10\,Gyr. This can be explained by the fact that the MSTO at this age becomes quite dim (according to PARSEC-COLIBRI isochrones $F475W=28.9$ at the distance of \object{M\,31}) and cluster luminosity is dominated by stars in the stochastically populated red giant branch (RGB).

The data presented in Fig.\,\ref{fig5} could be useful as a loose approximation of the completeness limit when analysing the \object{M\,31} PHAT \citep{Johnson2015} and the \object{M\,33} PHATTER \citep{Johnson2022} cluster catalogues; for example, studying the low-mass end of the cluster mass function. Furthermore, these estimates can be useful for more distant galaxies observed with HST \citep[e.g.][]{Scheepmaker2007, Ryon2015, Adamo2017, Cook2019, Deger2022}, by applying appropriate distance moduli, correcting for differing passbands, and carefully accounting for numerous systematic effects. Additionally, they could be helpful in planning future surveys with current or upcoming observatories as a point of reference to estimate the reachable survey depth.

A more qualitative presentation of stochastic sampling effects on cluster photometry is shown in Fig.\,\ref{fig6} and Fig.\,\ref{fig7}, where we plot two-colour diagrams (TCDs) and CMDs of simulated clusters, depending on the initial mass and age. In these figures, we also plot the PARSEC-COLIBRI non-stochastic simple stellar population (SSP) models of solar metallicity \citep{Marigo2017}. As was found in previous studies \citep{Deveikis2008, Fouesneau2010, deMeulenaer2013}, low-mass stochastic clusters (log$\rm(M/M_{\sun}) \leq 4.0$) exhibit large scatter, with CIs and fluxes spanning multiple magnitudes. Generally, the scatter of photometric parameters is stronger for younger and less massive clusters. This suggests that stochasticity can lead to large uncertainties and degeneracies \citep{Worthey1994, Bridzius2008} among the derived cluster age, extinction, metallicity, and mass. As can be seen in Fig.\,\ref{fig6} and Fig.\,\ref{fig7}, the brightest clusters in the cases of log$\rm(M/M_{\sun})=2.5$ and log$\rm(M/M_{\sun})=3.5$ show very similar colours and magnitudes in the visual -- near-infrared parts of spectra.

We notice that for low-mass star clusters younger than $\lesssim$1\,Gyr, due to the stochastic IMF sampling, two photometrically distinct groups of objects emerge (Fig.\,\ref{fig6} and Fig.\,\ref{fig7}). One group of clusters, which we call G1, are those that do not possess any PMS stars. Meanwhile, G2 clusters have at least one evolved PMS star. The distinction between these two types of clusters is particularly prominent at young ages ($\lesssim$100\,Myr), and it diminishes quickly at older ages, as the mass of stars at the MSTO decreases with age, and even low-mass clusters (log$\rm(M/M_{\sun}) \leq 3.0$) are populous enough to have some evolved stars.

Also, similarly to previous studies \citep{Deveikis2008, Bridzius2008, Fouesneau2010, Anders2013, deMeulenaer2013}, we demonstrate that in the case of low-mass star clusters, the non-stochastic SSP models, which are often used to derive physical parameters of stellar populations, are an inappropriate approximation. Figure\,\ref{fig6} and Fig.\,\ref{fig7} show that at young ages the non-stochastic SSP models lie in TCD and CMD areas poorly populated by stochastic clusters. The non-stochastic models poorly represent young clusters ($\leq$100\,Myr), where they lie in between stochastic cluster groups. It is noteworthy that at 1\,Gyr, non-stochastic SSP models match the distributions of stochastic clusters quite well. However, stochastic clusters still show a large scatter, covering wide ranges of non-stochastic model ages. Also, in TCDs ($F336W-F475W$ versus $F475W-F814W$), the natural direction of cluster spread is rather similar to the interstellar extinction vector. In general, these results further enforce the already established idea that use of non-stochastic SSP models to derive physical parameters of low-mass star clusters at certain ages is misleading, and stochasticity must be accounted for to obtain reliable parameter estimates.

Furthermore, we find that G1 clusters generally exhibit more consistent and robust photometric measurements, occupying a much smaller area of TCDs and CMDs, compared to G2 clusters. The latter are redder, cover a much broader range of CIs, and are brighter. It is important to stress that for low-mass clusters (log$\rm(M/M_{\sun})<4.0$), large photometric differences between two groups at young ages arise due to a handful, or in most cases, a single bright evolved PMS star, such as yellow or red supergiants (YSGs, RSGs), whose flux dominates over the remaining cluster. These findings are in agreement with \citet{Beerman2012} and suggest that more robust photometry of young low-mass star clusters can be obtained, and physical parameters can be derived with higher precision, if these bright evolved stars are somehow eliminated or avoided when performing aperture photometry of clusters.

\citet{Beerman2012} showed that by imposing a limiting magnitude and considering the flux below this limit -- in other words, eliminating bright evolved cluster stars from aperture photometry -- can provide more accurate age, mass, and extinction determinations, especially in the case of low-mass clusters. However, in such crowded fields as star clusters in galactic discs, it is an extremely difficult task to perform accurate stellar photometry, which would allow individual bright PMS stars to be subtracted, while maintaining the integrity of measurements for the remaining cluster. Nonetheless, the adaptive aperture photometry method, described in Sect.\,\ref{Sec3}, could be useful from this perspective as well. 

We would like to emphasise that clusters younger than 10\,Myr are generally dominated by the MS O-B stars, not RSGs or YSGs, which are still absent in majority of cases. Therefore, IMF-related stochasticity effects in these clusters are much weaker. However, these clusters were omitted from the present consideration because of more complex analysis required due to their location within natal star-forming clouds, which introduce differential extinction effects, and enhanced density of bright resolved field stars. 

Previous studies have shown that the cluster stochasticity problems can be restricted to certain parts of TCDs \citep{Whitmore2020} or to certain cluster ages (e.g. as already discussed, this occurs when RSGs, YSGs, or other bright PMS stars are present). Also, it is important to note that stochastic effects can be significantly reduced with the inclusion of ultraviolet, near-infrared \citep{Bridzius2008, deMeulenaer2014}, and H$\alpha$ \citep{Fouesneau2012, Whitmore2020, Whitmore2023} photometry data.

The adaptive aperture method was primarily designed to exclude bright field stars, which can introduce a strong bias in CIs, leading to older ages or larger extinctions being derived. However, PMS cluster members also introduce similar effects. Therefore, by excluding from measurements some of the evolved cluster stars residing further away from the cluster's centre, we can obtain more robust CIs and in turn derive more reliable cluster parameters. 

Although we successfully applied this method to the PHAT clusters \citep{Naujalis2021, Krisciunas2023} and obtained cluster colours that are more consistent and in better agreement with stochastic clusters, compared to the traditional aperture photometry approach, the adaptive aperture technique is yet to be tested extensively in controlled environments. Thus, in this study we applied the adaptive aperture method on background-less simulated cluster images to test the limits of such an approach, and most importantly, to investigate whether the assumption that there are no significant colour gradients arising due to stochastic effects is justified.

To determine whether any significant biases are introduced when colours are measured using  smaller colour apertures (C), compared to the conventional approach (T aperture), we define the difference $\Delta(CI)_{\rm TC}$ between these two measurements as
\begin{ceqn}
        \begin{align}
        \Delta(CI)_{\rm TC}=CI_{\rm T} - CI_{\rm C}, 
        \end{align}
\end{ceqn}
where $CI_{\rm T}$ is the colour index measured with T aperture; $CI_{\rm C}$ -- colour index measured with the C aperture. We show these differences for $F336W-F475W$ and $F475W-F814W$ CIs versus the radius of C aperture and cluster parameters in Fig.\,\ref{fig8} and Fig.\,\ref{fig9}.


For these CIs, we do not find any significant systematic colour gradients or biases, as the median of $\Delta(CI)_{\rm TC}$, regardless of the C aperture size and cluster parameters, remains close to 0. This supports the assumption made for the adaptive aperture photometry technique. Although there is no median colour bias, a rather significant spread of $\Delta(CI)_{\rm TC}$ exists. This scatter tends to decrease with an increasing C aperture size, as shown by error bars marking the 16th and 84th percentiles. Larger clusters (bigger $\rm r_{c}$, smaller $\gamma$) show a greater dispersion of $\Delta(CI)_{\rm TC}$ and it decreases more slowly with an increasing $\rm R_{C}$. This is understandable, as larger (less concentrated) clusters experience a higher level of stochasticity in their central parts, because smaller apertures capture only a small portion of the total cluster light. This trend with the C aperture size shows that the smaller the aperture is, the greater the uncertainty; therefore, it is generally beneficial to use a C aperture that is as large as possible. In \citet{Naujalis2021} and \citet{Krisciunas2023}, we mostly followed a rule that $\rm R_{C}$ has to be equal to or larger than the half-light radius, $\rm R_{50}$, of the corresponding cluster. 

In Fig.\,\ref{fig8} and Fig.\,\ref{fig9}, it can be seen that the scatter of $\Delta(CI)_{\rm TC}$ significantly decreases when aperture radii reach $\sim$2$\cdot \rm R_{50}$. For example, for clusters with $\rm r_{c}=0.1$\,arcsec and $\gamma=2.4$, or $\rm r_{c}=0.2$\,arcsec and $\gamma=2.8$, $\rm R_{50}$ is $\sim$0.4 and $\sim$0.5\,arcsec, respectively, meanwhile a strong decrease in $\Delta(CI)_{\rm TC}$ scatter, especially in $F475W-F814W$, can be observed at aperture radii of $\sim$0.8-1.0\,arcsec. Similarly, for more extended clusters ($\rm r_{c}=0.2$\,arcsec and $\gamma=2.4$, or $\rm r_{c}=0.4$\,arcsec and $\gamma=3.5$), when $\rm R_{50}$ is $\sim$0.7\,arcsec, a stronger decrease in the $\Delta(CI)_{\rm TC}$ scatter is observed at aperture radii of $\sim$1.2-1.5\,arcsec. Thus, a helpful orientation point when trying to choose an optimal aperture for colour measurements could be that its radius should be larger than $\rm R_{50}$, preferably $\sim$2$\cdot \rm R_{50}$, to minimise the effects of stochasticity. However, such a scheme may not yield the best results in every case, and we encourage users to try various apertures.

We notice that a number of clusters exhibit quite large $\Delta(CI)_{\rm TC}$ values, especially at the ages of 10 and 100\,Myr, some reaching $\pm$1.0\,mag or even more (see Fig.\,\ref{fig8} and Fig.\,\ref{fig9}). Most of these larger T and C colour measurement differences appear to be strong abrupt colour jumps. The majority of these sudden colour changes are positive, meaning that measurements with the T aperture provide redder CIs. At the age of 100\,Myr, $F475W-F814W$ appears to be more sensitive to these effects, compared to $F336W-F475W$, showing a larger dispersion of $\Delta(CI)_{\rm TC}$. This implies that the main origin of large differences between T and C aperture measurements is bright red PMS stars. To check if that is the case, before performing aperture photometry, we subtracted all PMS stars from simulated images. The results following such a procedure are shown on the right-hand panel blocks in Fig.\,\ref{fig8} and Fig.\,\ref{fig9}.

After the subtraction of PMS stars, the $\Delta(CI)_{\rm TC}$ scatter significantly diminishes, jumps of CIs mostly disappear, and all the differences between T and C measurements are around $\Delta(CI)_{\rm TC}=0$, mostly in the range of $\pm$0.1\,mag. An exception can be observed in the cases of a cluster age of log$\rm (t/yr)=7.0$ and masses of log$\rm(M/M_{\sun})\leq 3.0$. It can be seen that even after subtracting PMS stars, the $\Delta(CI)_{\rm TC}$ scatter remains rather strong, reaching around $\pm$0.5\,mag, and in some cases even more. Scatter is slightly biased towards negative values, meaning that C aperture measurements provide redder CIs. This shows that the stochasticity at the upper end of the MS for low-mass clusters can drive significant differences between measurements through T and C apertures in cases when the brightest MS stars are located outside the C aperture. Thus, users of the adaptive aperture method should be aware that for very young clusters the exclusion of the brightest MS stars by applying the C aperture can lead to determining redder CIs and in turn inaccurately derived physical parameters.

It is noteworthy that studies by \citet{Fleck2006} and \citet{Gaburov2008} show that if a strong mass segregation exists, significant colour gradients of 0.1-0.2\,mag can be observed in $V-I$ or $V-K$ for young ($<$100\,Myr) star clusters, and the use of the adaptive aperture method could give slightly biased results. However, \citet{Gaburov2008} estimated that in the case of log$\rm (t/yr)=7.0$, a minimum cluster mass of around log$\rm(M/M_{\sun})=5.7$ is necessary to distinguish radial colour gradients arising due to mass segregation and due to stochastic effects. Thus, for low-mass star clusters, adaptive aperture photometry assumptions are reasonable.

The adaptive aperture method enables users to obtain more accurate and consistent colour measurements, as in some cases a careful choice of C aperture helps to eliminate evolved cluster members that are outside of the cluster central part, whose inclusion in aperture photometry would increase the uncertainty of parameter derivation, as was shown by \citet{Beerman2012}. Also, this method  can be applied to obtain more accurate colour measurements in studies of low-mass clusters in the local Universe (up to distances of $\sim$5\,Mpc). With the resolution of HST photometry (and other present or future high-resolution missions), resolved bright field and evolved cluster stars can be excluded by a careful choice of C aperture. This method also can be applied to nearby galaxies, including \object{M\,31} \citep{Dalcanton2012, Johnson2015} and \object{M\,33} \citep{Williams2021, Johnson2022}, as well as to more distant galaxies observed in the vast HST surveys, such as LEGUS \citep{LEGUS, Adamo2017, Cook2019} or PHANGS-HST \citep{PHANGS, Deger2022}. Nonetheless, we would like to emphasise that the adaptive aperture method may not be a suitable or optimal approach for all star clusters or scientific goals, and we encourage users to try out various schemes to find the best solution for their particular cases.

\subsection{Accuracy of cluster aperture photometry}
\label{Sec42}

\begin{figure}
        \centering
        \includegraphics[width=9.0cm]{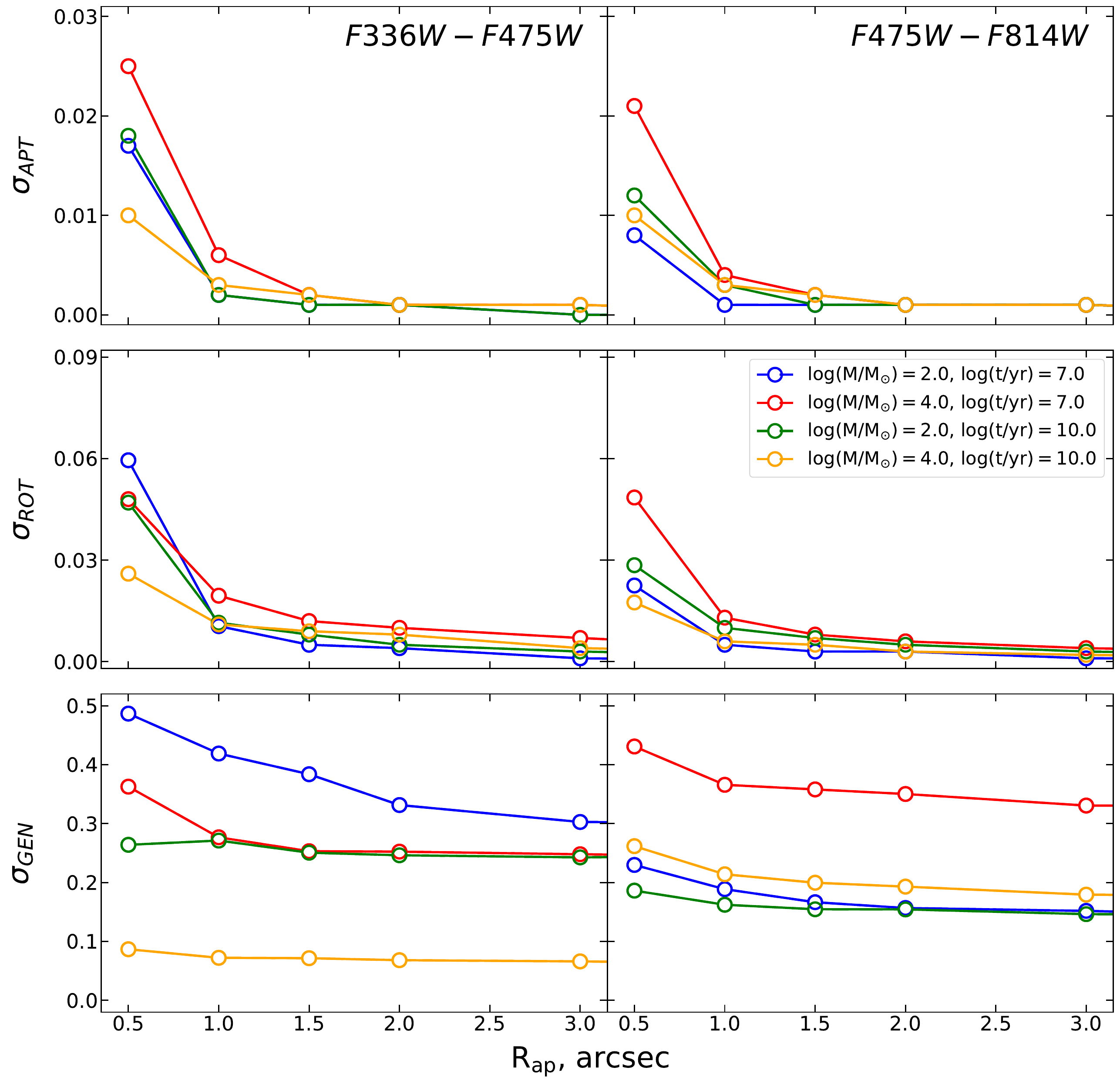}
        \caption{Star cluster aperture photometry errors and uncertainties of CIs $F336W-F475W$ and $F475W-F814W$ versus aperture radii. Top panels show the median of errors arising due to aperture positioning and size biases ($\sigma_{APT}$); middle panels -- the median of scatter due to projection effects ($\sigma_{ROT}$); bottom panels -- the median of scatter due to the stochastic IMF sampling and the spatial distribution of cluster stars, $\sigma_{GEN}$. Results for clusters with radial profile parameters, $\rm r_{c}=0.2$\,arcsec and $\gamma=2.8$, and 4 different cluster age and mass combinations (see legend) are shown.} 
        \label{fig10}
\end{figure}

Further on in this study, we attempted to quantify the intrinsic accuracy of the aperture photometry (Fig.\,\ref{fig10}). Statistical photometric errors that arise because of possible aperture centring and size biases were estimated in a similar fashion as in \citet{Naujalis2021} and \citet{Krisciunas2023}. We performed photometry at nine positions by taking nine central pixels (in the pixel scale of HST ACS/WFC cameras) of a simulated cluster image and centred the aperture on each of these pixels. At these positions, photometry by using an aperture of selected radius and two additional apertures with radii of $\pm$1\,pixel was performed. In total, this gave us 27 cluster magnitude and CI measurements of every simulated cluster projection (in total $3\cdot 10^7$ images) for a given aperture radius. We estimated possible uncertainties introduced by the cumulative aperture position and size biases as a half of the difference between the 84th and 16th percentile values, $\sigma_{APT}$. 

We simulated clusters by performing 100 independent stochastic IMF samplings and spatial distributions of cluster stars at each node of the model grid, and produced 100 individual clusters with the same parameters (age, mass, $\rm r_c$, $\gamma$).  Scatters of cluster photometric parameters, arising due to stochastic IMF sampling and spatial distribution of cluster stars (GEN), were estimated as a half of the differences between the 84th and 16th percentile values, $\sigma_{GEN}$.

Since we simulated images of every cluster mimicking their observations from 100 different directions, this enabled us to evaluate the projection effects on photometric measurements. Scatters of cluster photometric parameters, arising due to projection effects (ROT), were estimated as a half of the differences between the 84th and 16th percentile values, $\sigma_{ROT}$. 

In Fig.\,\ref{fig10}, we show uncertainties, arising due to these three sources, for CIs $F336W-F475W$ and $F475W-F814W$ in the cases of four different combinations of cluster age and mass (log$\rm (t/yr)=7.0$ and 10.0; log$\rm(M/M_{\sun})=2.0$ and 4.0) versus aperture radii. The radial profile parameters consistent with typical clusters observed in \object{M\,31} ($\rm r_{c}=0.2$\,arcsec and $\gamma=2.8$) are assumed. It can be seen that in the cases of all three uncertainty sources, the steepest decrease in uncertainties is in the aperture size range from $\rm R_{ap}=0.5$ to 1.0\,arcsec. All these uncertainty sources more or less stabilise at $\rm R_{ap}\sim 1.5$\,arcsec. When apertures of $\rm R_{ap}\geq 1.5$\,arcsec are used, $\sigma_{ROT}$ becomes insignificant -- below $\sim$0.01\,mag. Nonetheless, in our studies of the PHAT clusters \citep{Naujalis2021, Krisciunas2023}, in most cases, radii of C apertures are smaller than 1.5\,arcsec; thus, ROT effects can limit the highest achievable accuracy of photometry.

However, the main takeaway from the results presented in Fig.\,\ref{fig10} is that GEN and ROT effects (especially GEN, which generally is several times larger than ROT) are much stronger than the ones due to the aperture positioning and size biases. This means that in conditions when the background noise and field stars are perfectly accounted for, IMF stochasticity is the main factor limiting the accuracy of star cluster aperture photometry. Nonetheless, readers should be aware that in star cluster studies, besides the random uncertainty sources reported here, some systematic issues exist; for example, the degeneracy between various physical parameters \citep{Worthey1994, Bridzius2008, Whitmore2023}. In some particular cases, such as old globular clusters, these systematic effects can be a much stronger limiting factors on our ability to derive accurate cluster parameters than the random photometry errors introduced by stochasticity.

We would like to emphasise, based on the results provided in Fig.\,\ref{fig10}, that it is important to consider uncertainties of cluster measurements introduced by the randomness of projection towards the line of sight in the case of extragalactic star clusters. We have found that this uncertainty essentially surpasses systematic errors of aperture photometry procedures and must be taken into the account when considering the total error budget of star cluster measurements. In the case of typical size clusters located at the distance of \object{M\,31} and measured with apertures of $\rm R_{ap}\leq 1.5$\,arcsec, ROT effects can cause uncertainties of CIs up to $\sim$0.1\,mag, depending on cluster parameters and the presence of PMS stars. More massive and more compact clusters exhibit a lower $\sigma_{ROT}$, especially if PMS stars are absent. In any case, the use of bigger apertures minimises the impact of projection effects. However, it is important to stress that use of larger apertures can reduce the overall quality of the photometry. Thus, the optimal approach might be different for each particular cluster and a specific research task.

\subsection{Physical cluster parameters}
\label{Sec43}

\begin{figure}
        \centering
        \includegraphics[height=10.5cm]{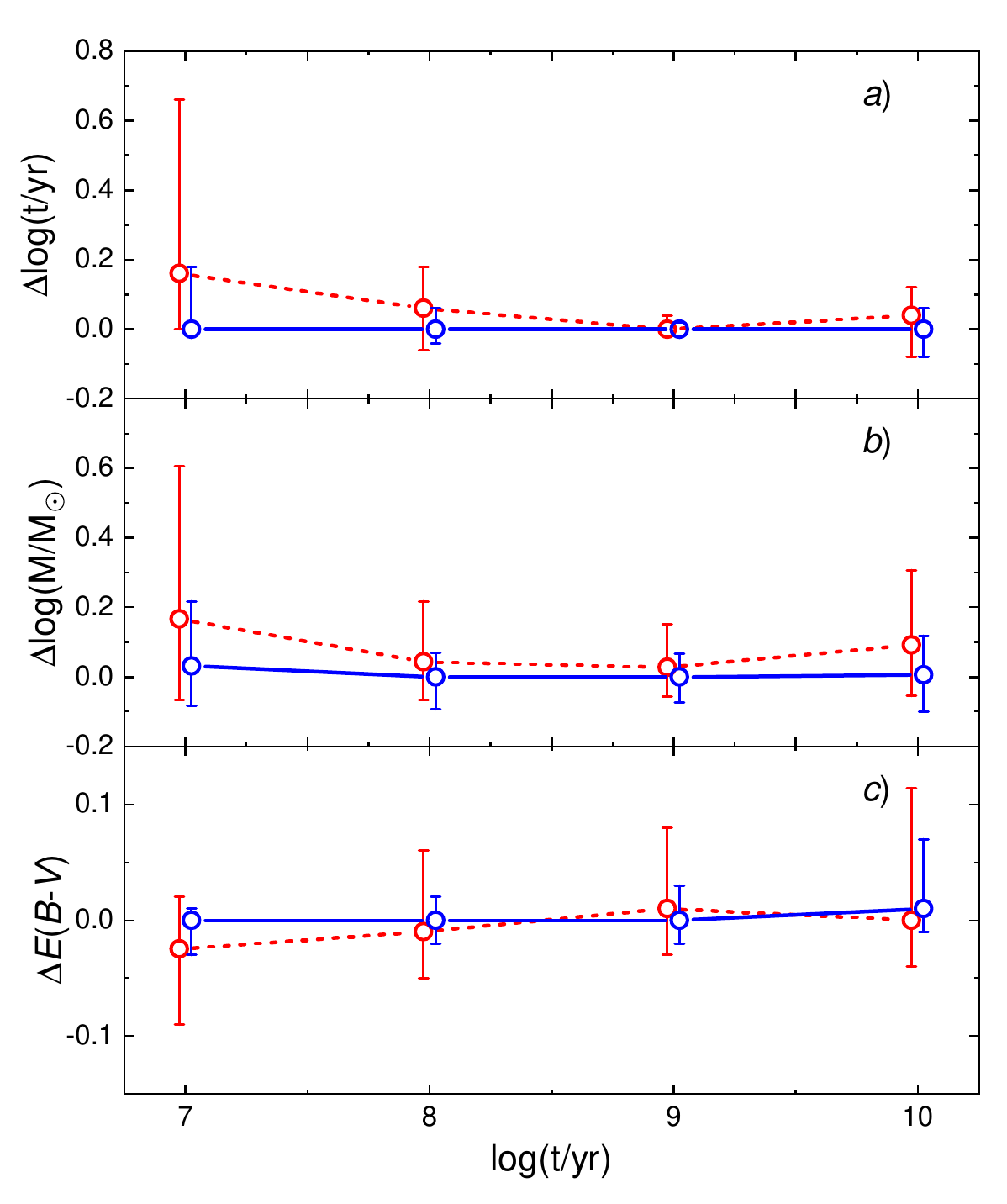}
        \caption{Results of star cluster physical parameter determination tests. The provided differences are between the determined and input parameter values. Panel a) shows the median of age differences; b) -- the median of mass differences; c) -- the median of colour excess, $E(B-V)$, differences. Red open circles mark median values of parameter differences determined based on the measurements through the aperture of $\rm R_{ap}=0.5$\,arcsec. Blue open circles mark median values of parameter differences determined based on the measurements through the apertures of $\rm R_{ap}=1.0$, 2.0, and 7.0\,arcsec, and then merged into a single dataset. Error bars show the 16-84 percentile ranges. Markers and corresponding lines are slightly shifted along the X-axis around the original input value for visual clarity.} 
        \label{fig11}
\end{figure}

To estimate the accuracy limits of cluster physical parameter derivation -- age, mass, and extinction, $E(B-V)$ -- based on the aperture photometry results, we employ the photometry of modelled cluster images. For the results presented in this section, we used artificial clusters with log(t/yr)\,=\,[7.0, 8.0, 9.0, 10.0]; log$\rm(M/M_{\sun})$\,=\,[2.5, 3.0, 3.5, 4.0]; $\gamma$\,=\,[2.2, 2.8, 7.0]; $\rm r_{c}$\,=\,[0.1, 0.4]\,arcsec. At each parameter node, we analysed 20 different clusters. Therefore, we analysed 1920 different clusters in total. 

For the parameter determination tests, we used photometric measurements of artificial clusters in six PHAT passbands, and by applying four apertures of the following sizes: $\rm R_{ap}$\,=\,[0.5, 1.0, 2.0, 7.0]\,arcsec. All 1920 simulated clusters were measured through all these apertures. The determination of cluster parameters (mass, age, extinction) was performed by comparing aperture photometry measurements with the large grid of stochastic cluster models. For this purpose, we used the method described in \citet{deMeulenaer2013, deMeulenaer2017}.

The cluster model grid used for parameter determination contains the following nodes: age from log$\rm (t/yr)=6.0$ to 10.3 in steps of 0.02; mass from log$\rm(M/M_{\sun})=2.0$ to 7.0 in steps of 0.05; extinction from $E(B-V)=-0.03$ to 1.5 in steps of 0.01, totalling to $\sim$$4\cdot10^{6}$ nodes. Each node of the grid contains 10\,000 stochastic clusters. Therefore, in total there are $\sim$$4\cdot10^{10}$ solar metallicity cluster models. Parameter probabilities for each of the 1920 test clusters measured through four apertures were calculated following Eq.\,1 and 2 from \citet{deMeulenaer2017}. We selected the maximum likelihood parameter combinations.

Accuracies of the derived parameters were estimated by calculating the differences between the determined and input parameter values: $\Delta \rm log(t/yr)=\rm log(t/yr)-\rm log(t/yr)_{input}$, $\Delta \rm log(M/M_{\sun})=\rm log(M/M_{\sun})-\rm log(M/M_{\sun})_{input}$, $\Delta E(B-V)=E(B-V)-E(B-V)_{\rm input}$. Initially, we used two $E(B-V)_{\rm input}$ values -- 0.0 and 0.5; however, the differences of derived parameter values were negligible; thus, further in this paper we discuss only the results obtained using $E(B-V)_{\rm input}=0.0$.

The main results of cluster parameter determination tests are presented in Fig.\,\ref{fig11}. We show summarised medians of parameter differences versus cluster ages. We did not find significant differences in scatter and median values of $\Delta \rm log(t/yr)$, $\Delta \rm log(M/M_{\sun})$, and $\Delta E(B-V)$ when using apertures of radii $\rm R_{ap}=1.0$, 2.0, and 7.0\,arcsec. Also, there were no significant dependencies of $\Delta \rm log(t/yr)$, $\Delta \rm log(M/M_{\sun})$, and $\Delta E(B-V)$ distributions on $\rm r_{c}$ and $\gamma$. Therefore, we accumulated parameter differences derived based on the measurements through those three apertures and then merged them into one dataset. It is represented by blue open circles and lines in Fig.\,\ref{fig11}. Here, each data point represents 480 artificial star clusters measured through three apertures; that is, 1440 parameter differences. Meanwhile, parameter differences derived based on the measurements through the aperture of radius $\rm R_{ap}=0.5$\,arcsec are significant, compared to other apertures. Therefore, we show these results by red open circles and dashed lines. In this case, each data point represents 480 artificial star clusters. Error bars represent the 16-84 percentile range of the respective parameter difference distributions.

One of the most important features observed in Fig.\,\ref{fig11} is the lower accuracy of parameters  in the case of $\rm R_{ap}=0.5$\,arcsec, compared to the results based on measurements with larger apertures. Also, biases of median values of parameter distributions are significant. It is noteworthy that the median values of parameter distributions, based on measurements with larger apertures, are unbiased with high precision and demonstrate a much lower spread of the derived parameters. 

At 10\,Myr, the medians of $\Delta \rm log(t/yr)$ and $\Delta \rm log(M/M_{\sun})$ are shifted by around +0.2\,dex, implying older ages and bigger masses being derived. Such a correlation is not surprising -- if an older age is derived, then a higher mass is needed to match the integrated flux. This suggests that using $\rm R_{ap}=0.5$\,arcsec or smaller apertures can introduce additional parameter degeneracies. In general, it is clear that use of apertures with radii of $\rm R_{ap}\leq 0.5$\,arcsec is not advisable throughout the entire parameter space.

These findings are in agreement with the restrictions to the adaptive aperture method established in \citet{Naujalis2021} and \citet{Krisciunas2023}, and were already discussed in Sect.\,\ref{Sec41}. Radii of the C apertures, dedicated for measuring cluster CIs, have to be larger than $\rm R_{50}$, since a typical cluster size in \object{M\,31} is $\rm R_{50}\sim 0.5$\,arcsec. Following this rule enabled us to obtain robust cluster photometric measurements, and in turn, more accurately derived physical parameters.

However, it is important to note that parameter determination for clusters of 10\,Myr age  encounters serious problems, regardless of the aperture size (even in the case of $\rm R_{ap}=7.0$\,arcsec aperture, which measures all of the cluster's flux). At this age, the 16-84 percentile range of $\Delta \rm log(t/yr)$ is almost 0.7\,dex when measured with $\rm R_{ap}=0.5$\,arcsec and $\sim$0.2\,dex with larger apertures, respectively. At 10\,Myr, the 16-84 percentile range of $\Delta \rm log(M/M_{\sun})$ also spans $\sim$0.7\,dex with $\rm R_{ap}=0.5$\,arcsec and $\sim$0.3\,dex when $\rm R_{ap}\geq 1.0$\,arcsec, respectively. The behaviour of $\Delta E(B-V)$ is different: the lowest accuracy is apparent in the case of old (10\,Gyr) clusters. 

In summary, it is evident that the derivation of physical parameters of young star clusters, based on aperture photometry, can lead to low accuracies, as established in Fig.\,\ref{fig6} and Fig.\,\ref{fig7}. Moreover, considering that uncertainties reported in this study represent idealised cases (without the sky background); in realistic environments the accuracies would be even lower. Therefore, the parameters of resolved and semi-resolved young star clusters ($\lesssim$30\,Myr) determined using CMD analysis methods \citep{Weisz2015, Johnson2016, Wainer2022, Ceponis2024} would most likely yield more accurate results compared to the ones based on aperture photometry techniques \citep{Deveikis2008, Fouesneau2010, deMeulenaer2013, Fouesneau2014, Krumholz2015, deMeulenaer2017}. Also, it is important to stress that in addition to the stochastic effects studied in this paper, the accuracy of derived physical cluster parameters  depends on the stellar evolution and population synthesis models used, as the differences between various models are still significant.

\subsection{Structural cluster parameters}
\label{Sec44}

\begin{figure}
        \centering
        \includegraphics[width=9cm]{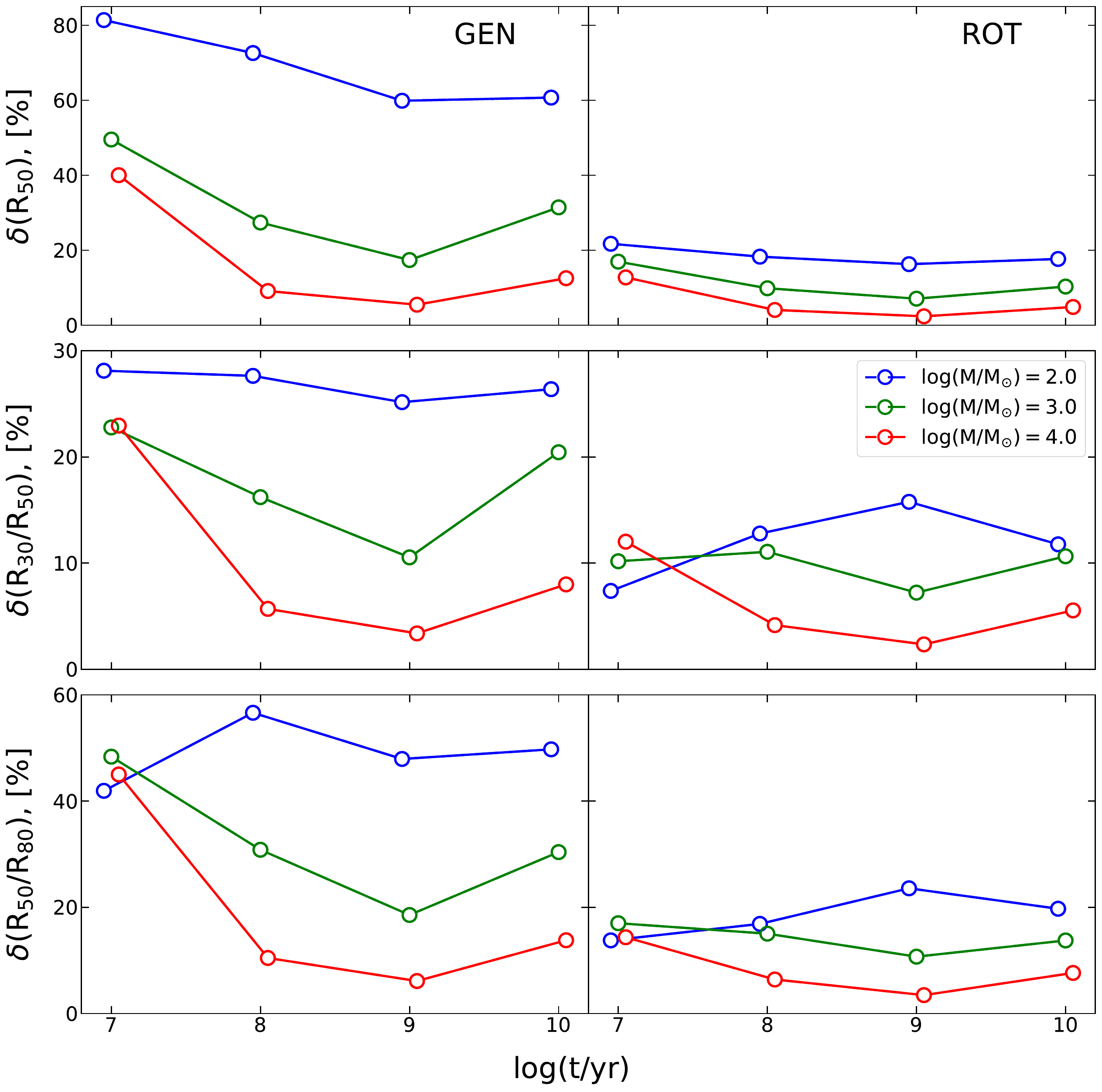}
        \caption{Uncertainties, expressed in percent, of star cluster structural parameters versus cluster age. Left column -- the median of uncertainties arising due to stochastic IMF sampling and the spatial distribution of cluster stars, $\delta_{GEN}$; right column --  due to cluster projection effects, $\delta_{ROT}$. Top row -- the median of uncertainties of $\rm R_{50}$; middle row -- $\rm R_{30}/R_{50}$; bottom row -- $\rm R_{50}/R_{80}$. Results are shown for clusters with $\rm r_{c}=0.2$\,arcsec, $\gamma=2.8$, and three mass values (see legend). Structural parameters are measured using FGCs in the $F475W$ passband. Blue and red markers and corresponding lines are slightly shifted along the X-axis around the original input value for visual clarity.} 
        \label{fig12}
\end{figure}
\begin{figure}
        \centering
        \includegraphics[width=9cm]{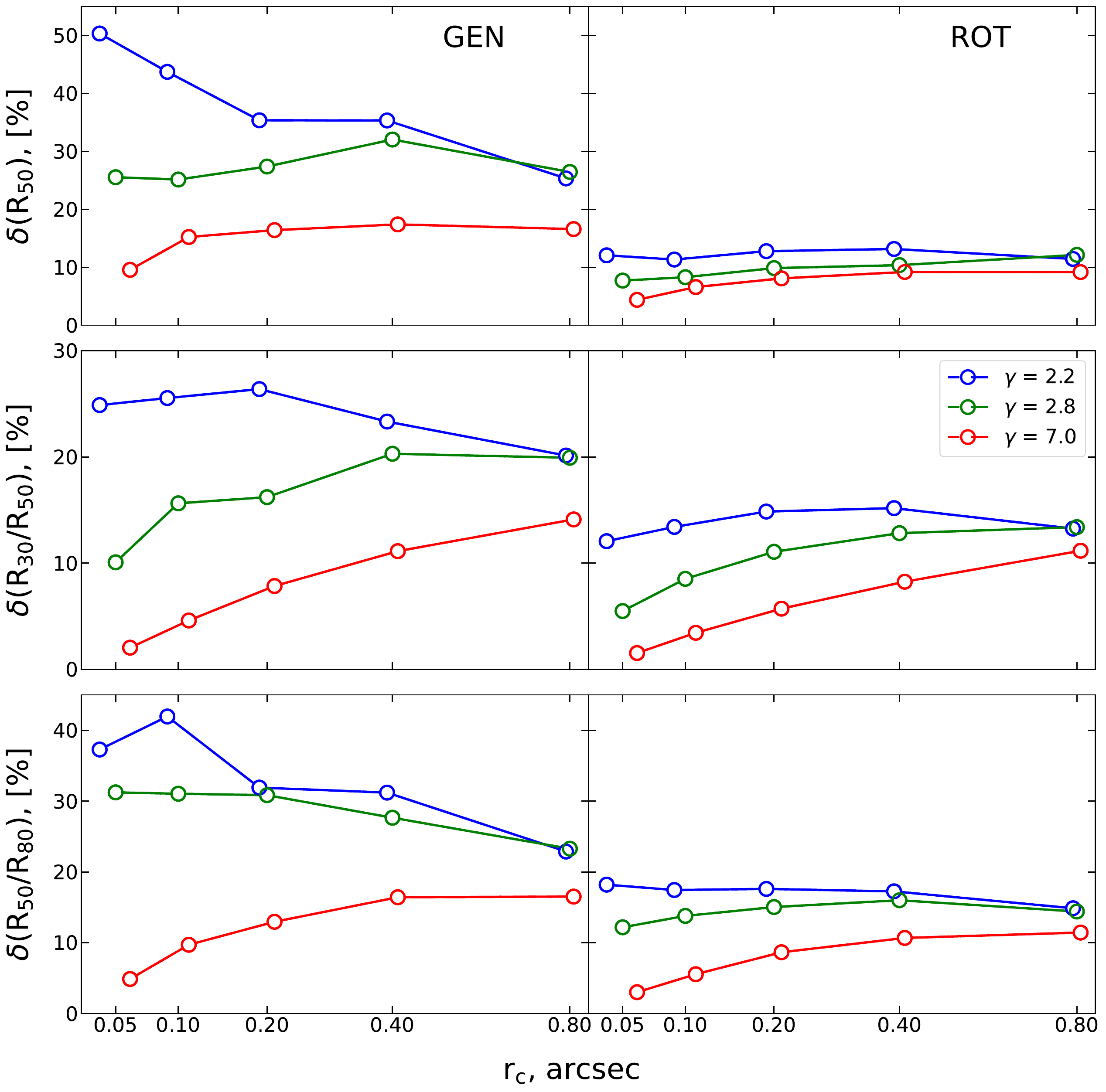}
        \caption{Uncertainties, expressed in percent, of star cluster structural parameters versus cluster core radius, $\rm r_{c}$. Left column -- the median of uncertainties arising due to stochastic IMF sampling and the spatial distribution of cluster stars, $\delta_{GEN}$; right column -- due to cluster projection effects, $\delta_{ROT}$. Top row -- the median of uncertainties of $\rm R_{50}$; middle row -- $\rm R_{30}/R_{50}$; bottom row -- $\rm R_{50}/R_{80}$. Results are shown for clusters with log$\rm(M/M_{\sun})=3.0$ and log$\rm (t/yr)=8.0$, and three $\gamma$ values (see legend). Structural parameters are measured using FGCs in the $F475W$ passband. Blue and red markers and corresponding lines are slightly shifted along the X-axis around the original input value for visual clarity. Square root scaling is applied for the X-axis.}
        \label{fig13}
\end{figure}
\begin{figure}
        \centering
        \includegraphics[width=9cm]{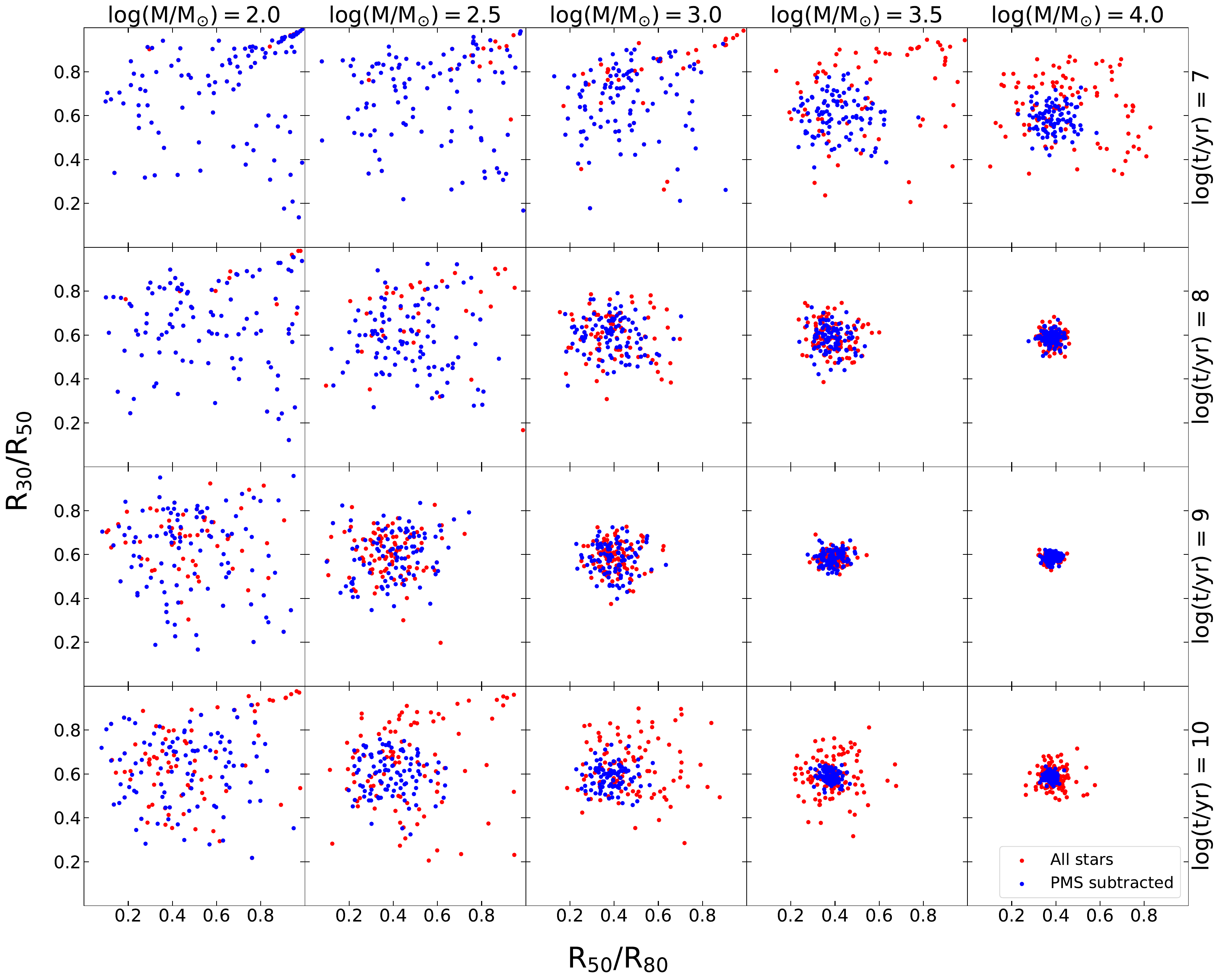}
        \caption{Effects of stochastic IMF sampling and the spatial distribution of stars on cluster structural parameters depending on cluster mass and age. The X-axis represents $\rm R_{50}/R_{80}$; the Y-axis -- $\rm R_{30}/R_{50}$. Results are shown for clusters with $\rm r_{c}=0.2$\,arcsec and $\gamma=2.8$. Structural parameters are measured in the $F475W$ passband. Clusters that possess all stars are marked with red dots, while clusters with PMS stars subtracted -- with blue dots.}
        \label{fig14}
\end{figure}
\begin{figure}
        \centering
        \includegraphics[width=9cm]{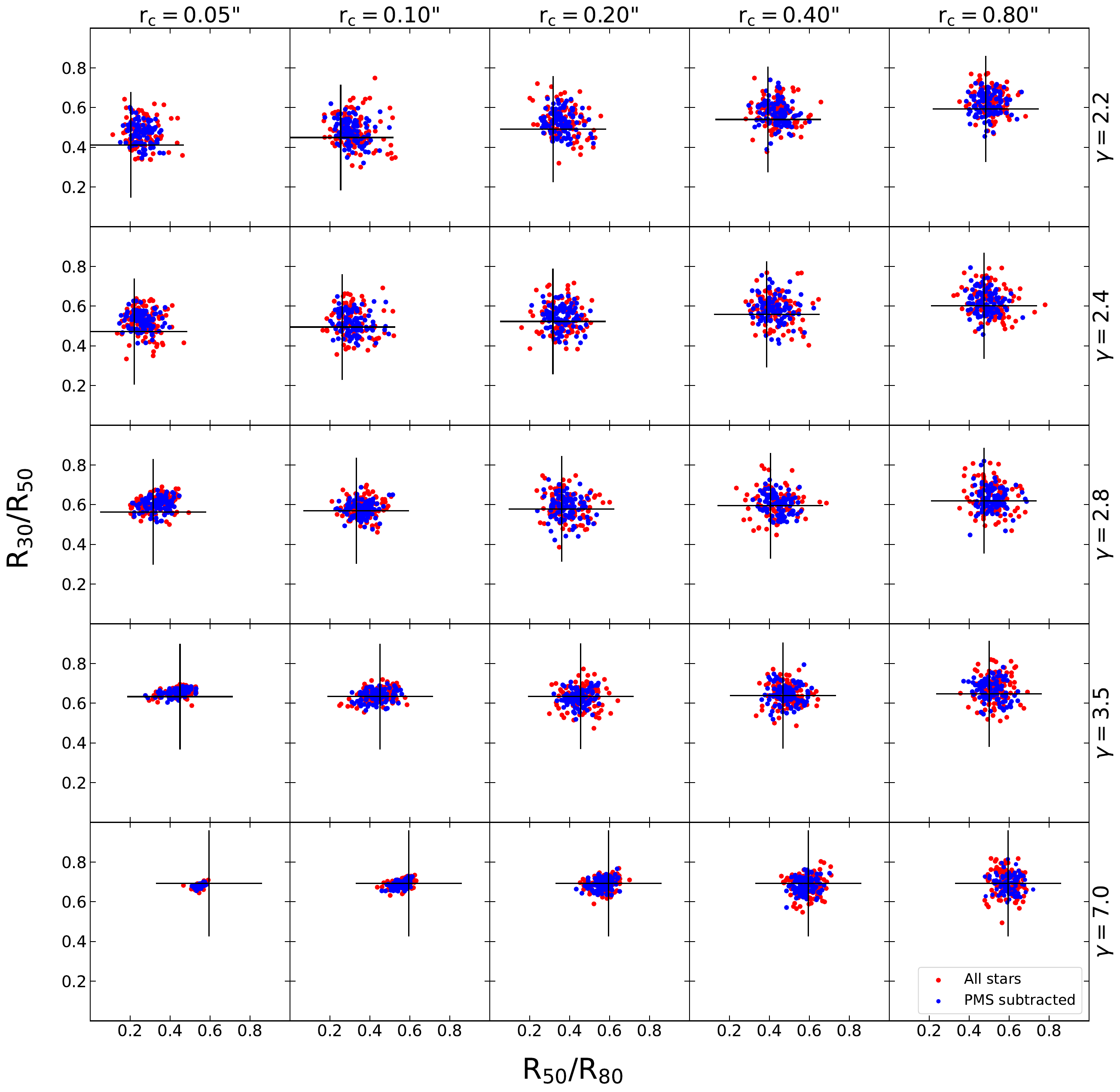}
        \caption{Effects of stochastic IMF sampling and the spatial distribution of stars on cluster structural parameters depending on cluster radial profile parameters, $\rm r_{c}$ and $\gamma$. The X-axis represents $\rm R_{50}/R_{80}$; the Y-axis -- $\rm R_{30}/R_{50}$. Results are shown for clusters with log$\rm(M/M_{\sun})=3.5$ and log$\rm (t/yr)=8.0$. Structural parameters are measured in the $F475W$ passband. Clusters that possess all stars are marked with red dots, while clusters with PMS stars subtracted -- with blue dots. The crosshairs mark the theoretical locations of star clusters calculated using input radial profile parameters.}
        \label{fig15}
\end{figure}

In this section, we investigate how the stochastic nature of low-mass star clusters affects uncertainties of FGC-based cluster structural parameters. We define the scatters of structural parameters that arise due to IMF sampling and the spatial distribution of cluster stars ($\sigma_{GEN}$), as well as projection effects ($\sigma_{ROT}$), in the same manner as we did for cluster photometric parameters (see Sect.\,\ref{Sec42}): as a half of the differences between the 84th and 16th percentile values. However, since R parameters and their ratios are strongly dependent on the input EFF parameters, we investigate the relative uncertainties ($\delta_{GEN}$ and $\delta_{ROT}$) in terms of percentage, which are calculated as the ratios of structural parameter scatter measures ($\sigma_{GEN}$ and $\sigma_{ROT}$) and median values of the corresponding parameters.

In Fig.\,\ref{fig12}, we show the uncertainties, arising due to GEN and ROT effects, for $\rm R_{50}$, $\rm R_{30}$/$\rm R_{50}$, and $\rm R_{50}$/$\rm R_{80}$, in the cases of three cluster masses (log$\rm(M/M_{\sun})=2.0$, 3.0, and 4.0) versus cluster ages. Here, we use EFF profile parameters characteristic to young clusters observed in \object{M\,31}, $\rm r_{c}=0.2$\,arcsec and $\gamma=2.8$ \citep{Sableviciute2007, Vansevicius2009}. Due to GEN effects, the cluster size estimate, $\rm R_{50}$, varies strongly, even though cluster input parameters are fixed. It is especially prominent in the case of very young objects -- at the age of 10\,Myr throughout the entire mass interval studied in this paper, $\delta_{GEN}\gtrsim40$\%. Also, GEN effects introduce a $\rm R_{50}$ scatter of $\sim$20\% in the studied age range for clusters with log$\rm(M/M_{\sun})\leq 3.0$. Only massive clusters with log$\rm(M/M_{\sun})=4.0$ at the ages of 100\,Myr and 1\,Gyr show $\rm R_{50}$ uncertainties, arising due to GEN effects, that are smaller than 10\%.

Additionally, ROT effects impose some accuracy limits on FGC-based estimates of $\rm R_{50}$.  Throughout the entire studied age and mass intervals, uncertainties of $\rm R_{50}$, arising due to ROT effects, are smaller than those due to GEN effects for corresponding clusters. At the ages of log$\rm (t/yr) \geq 8.0$, the clusters of log$\rm(M/M_{\sun})\geq 3.0$ mass exhibit $\delta_{ROT}<10$\%.

Considering R parameter ratios -- $\rm R_{30}$/$\rm R_{50}$ and $\rm R_{50}$/$\rm R_{80}$ -- throughout the entire studied cluster parameter ranges, both GEN and ROT effects are smaller for $\rm R_{30}$/$\rm R_{50}$. GEN effects for clusters with log$\rm(M/M_{\sun})\leq 3.0$ cause $\sim$10-30\% uncertainties of $\rm R_{30}$/$\rm R_{50}$, and $\sim$20-60\% of $\rm R_{50}$/$\rm R_{80}$. Meanwhile, in the same mass interval, ROT effects impose a typical $\rm R_{30}$/$\rm R_{50}$ uncertainty of 5-15\%, and 10-25\% in the case of $\rm R_{50}$/$\rm R_{80}$. However, at 10\,Myr, throughout the entire studied cluster mass range, $\delta_{GEN}$ is $\gtrsim$20\% and $\gtrsim$40\% for $\rm R_{30}$/$\rm R_{50}$ and $\rm R_{50}$/$\rm R_{80}$, respectively. Meanwhile, uncertainties arising due to ROT effects at this age are $\sim$10\% and $\sim$15\% for $\rm R_{30}$/$\rm R_{50}$ and $\rm R_{50}$/$\rm R_{80}$, respectively. Once again, massive (log$\rm(M/M_{\sun})=4.0$) clusters of an age log$\rm (t/yr) \geq 8.0$ exhibit more accurate $\rm R_{30}$/$\rm R_{50}$ and $\rm R_{50}$/$\rm R_{80}$ structural parameters in terms of uncertainties arising due to GEN and ROT effects. In general, the ratios $\rm R_{30}$/$\rm R_{50}$ are more robust estimates of cluster structures than $\rm R_{50}$/$\rm R_{80}$. This implies that in most cases, the central parts of clusters are less susceptible to stochastic effects than the outer regions.

Structural parameters based on FGCs are generally more robust for higher mass clusters, with the exception of the youngest objects (Fig.\,\ref{fig12}). Structural parameter uncertainties, arising due to GEN and ROT effects, tend to decrease with age up to 1\,Gyr and then increase at 10\,Gyr. However, clusters with a mass of log$\rm(M/M_{\sun})=2.0$ exhibit an inverse behaviour of $\delta_{ROT}$ to the aforementioned age trend in the cases of $\rm R_{30}$/$\rm R_{50}$ and $\rm R_{50}$/$\rm R_{80}$. That is because such low-mass clusters are unlikely to have PMS stars at ages of 10\,Myr. However, as the mass of stars at MSTO decreases with age, these clusters start to contain dominant PMS stars. The evolved stars increase the uncertainty of structural parameters as the shapes of clusters' FGCs strongly depend on the location of the PMS stars.

In Fig.\,\ref{fig13}, we show uncertainties, arising due to GEN and ROT effects, for $\rm R_{50}$, $\rm R_{30}$/$\rm R_{50}$, and $\rm R_{50}$/$\rm R_{80}$, in the cases of three $\gamma$ parameters ($\gamma=2.2$, 2.8, and 7.0) versus cluster core radii, $\rm r_{c}$. Here, cluster models with log$\rm(M/M_{\sun})=3.0$ and log$\rm (t/yr)=8.0$ are assumed to represent typical clusters in \object{M\,31}. General trends for both uncertainty sources are such that smaller (more concentrated) clusters experience a lower uncertainty of structural parameters. However, objects with $\gamma=2.2$ are a noteworthy exception; it can be seen that for these clusters, uncertainties in percent, especially due to GEN effects, decrease as $\rm r_{c}$ increases. It can be noted that, in the case of $\rm R_{50}$/$\rm R_{80}$, even $\gamma=2.8$ clusters show a similar behaviour. We find that this phenomenon occurs because in those cases, even though the  absolute uncertainties ($\sigma_{GEN}$ and $\sigma_{ROT}$) grow with an increasing $\rm r_{c}$, the median values of the respective parameters increase at a faster rate. Thus, this leads to a percentage decrease in uncertainties ($\delta_{GEN}$ and $\delta_{ROT}$) at larger core radii, $\rm r_{c}$.

We investigated how stochastic IMF sampling and the spatial distribution of stars affects radial profiles of star clusters by exploring their behaviour in the parameter space of $\rm R_{30}$/$\rm R_{50}$ versus $\rm R_{50}$/$\rm R_{80}$. In Fig.\,\ref{fig14}, we show distributions of clusters in this parameter space depending on their mass and age. Unfortunately, in many cases of cluster mass and age combinations, there is a large scatter -- clusters cover most of the parameter space without a clear concentration. This feature is prominent for clusters with log$\rm(M/M_{\sun})\leq 2.5$ throughout the entire age range. A similar situation is observed for young (10\,Myr) clusters when PMS stars are not subtracted (red dots). 

An important aspect should be noted: for clusters of the lowest mass (log$\rm(M/M_{\sun})=2.0$) at all ages, and for clusters of moderate masses (log$\rm(M/M_{\sun})\leq 3.0$) at an age of log$\rm (t/yr)=7.0$, there is a significant number of clusters exhibiting both R parameter ratios close to 1.0 in the parameter space of $\rm R_{30}$/$\rm R_{50}$ versus $\rm R_{50}$/$\rm R_{80}$. This implies that a single star dominates the clusters' FGCs, as $\rm R_{30}$, $\rm R_{50}$, and $\rm R_{80}$ are almost equal.  

We find that star clusters with masses of log$\rm(M/M_{\sun})\geq 3.5$ and ages of log$\rm (t/yr)\geq 8.0$ exhibit rather well-defined and concentrated distributions in the parameter space $\rm R_{30}$/$\rm R_{50}$ versus $\rm R_{50}$/$\rm R_{80}$; this is especially prominent for the clusters with a mass of log$\rm(M/M_{\sun})=4.0$ and ages of log$\rm (t/yr) \geq 8.0$. This implies that aperture-photometry-based structural parameters for such clusters can be estimated with a reasonable accuracy. It is worth mentioning that clusters of log$\rm(M/M_{\sun})\geq 3.5$ at the age of log$\rm (t/yr)=7.0$ show a significant improvement when PMS stars are subtracted (blue dots). Yet, log$\rm(M/M_{\sun})=4.0$ stochastic clusters in most cases at an age of log$\rm (t/yr)=7.0$ have multiple bright evolved stars; therefore, in realistic environments, it could be quite difficult to subtract all of them, as some of PMS stars can be located in the unresolved central part of the cluster. On the other hand, subtracting only a part of the bright stars could lead to additional uncertainties and inconsistencies. 

In Fig.\,\ref{fig15}, we show the behaviour of clusters in the parameter space of $\rm R_{30}$/$\rm R_{50}$ versus $\rm R_{50}$/$\rm R_{80}$ depending on radial profile parameters, $\rm r_{c}$ and $\gamma$. The results for simulated clusters with log$\rm(M/M_{\sun})=3.5$ and log$\rm (t/yr)=8.0$ are shown. Most compact clusters ($\gamma=3.5$-7.0, $\rm r_{c}=0.05$-0.10\,arcsec) exhibit a small scattering of structural parameters. Even though the general trend is that more extended clusters experience slightly stronger scatter, visually its level is quite similar when $\gamma\leq 2.8$, $\rm r_{c}\geq 0.20$\,arcsec. It is worth noting that clusters gather around slightly different points of the R parameter space, depending on the values of $\rm r_{c}$ and $\gamma$. This suggests that by determining clusters' locations in the parameter space of $\rm R_{30}$/$\rm R_{50}$ versus $\rm R_{50}$/$\rm R_{80}$ we can obtain useful information on their structure, specifically in the cases of massive star clusters, log$\rm(M/M_{\sun}) > 3.0$. At lower masses, the scatter is too large and clusters with different $\rm r_{c}$ and $\gamma$ values overlap in the parameter space of $\rm R_{30}$/$\rm R_{50}$ versus $\rm R_{50}$/$\rm R_{80}$.

Overall, the results discussed in this section imply that structural parameters for low-mass (log$\rm(M/M_{\sun}) < 3.5$) young (log$\rm (t/yr) < 8.0$) star clusters, derived solely based on FGCs, are rather uncertain and should be interpreted cautiously. The randomness of cluster projection towards the line of sight (ROT) introduces uncertainties of 10-20\% in cluster size ($\rm R_{50}$) estimates. However, the scatter of $\rm R_{50}$ values arising due to stochastic IMF sampling and the spatial distribution of cluster stars (GEN) is even larger. \citet{Narbutis2014} found that accounting for bright resolved stars significantly improves the accuracy of derived structural parameters when a 2D cluster image fitting is performed. However, in our case, the accuracy improvement of the FGC-based structural parameters by subtracting bright evolved stars does not solve problems for all cluster parameter combinations. 

Considering low-mass star clusters, to investigate and draw reliable conclusions on the cluster mass-radius relation, cluster disruption, their dynamical evolution \citep{PortegiesZwart2010, Krumholz2019, Adamo2020, Brown2021}, aperture-photometry-based structural parameters are insufficient. Therefore, additional complementary methods should be applied; for example, the fitting of the 2D cluster images \citep{Larsen1999, Narbutis2014, Brown2021}.

\section{Summary and conclusions}
\label{Sec5}

In this study, we determine the intrinsic maximal achievable accuracy and applicability limits of the aperture photometry method for star cluster studies in the local Universe. To do so, we simulated a large grid of 3D cluster models ($5\cdot 10^4$) covering the parameter space of real clusters observed in \object{M\,31}. Images of these artificial clusters were modelled by projecting each of them onto a 2D plane from differing viewing angles, mimicking the observation of the same object from 100 directions. Images were generated in six passbands (making a total of $3\cdot 10^7$ images) in a manner consistent with the PHAT survey \citep{Dalcanton2012}. To investigate the accuracy and applicability limits of the aperture photometry for star clusters, we measured modelled images and ran parameter determination tests.

We show that groups of clusters with (G2) and without (G1) PMS stars photometrically differ significantly; the scatter of photometric parameters and measurement uncertainties are much larger for G2 objects. Our findings are in agreement with \citet{Beerman2012} -- the absence of PMS stars results in much more consistent and well-defined cluster photometric parameters, which in turn lead to a more robust physical parameter derivation.

We find that there are no significant systematic differences in CIs when star clusters are measured using smaller (C) apertures and larger, conventional (T) apertures. This justifies the adaptive aperture photometry method introduced by \citet{Naujalis2021} to minimise the contamination of CI measurements with bright field and evolved cluster stars. We suggest that this approach with a careful choice of C apertures can be applied to obtain more accurate photometric measurements of low-mass star clusters in the local Universe ($\lesssim$5\,Mpc).

As is shown by the results of the physical parameter determination tests, to derive accurate and reliable aperture-photometry-based star cluster parameter estimates, 
it can be beneficial to measure CIs using apertures with radii that are larger than the clusters' half-light radii, as is suggested by \citet{Naujalis2021}. However, we find that parameter determination of very young star clusters ($\sim$10\,Myr) is problematic, regardless of the aperture size used. Thus, using CMD fitting methods \citep{Weisz2015, Johnson2016, Wainer2022, Ceponis2024} instead would most likely enable the determination of more reliable physical parameters of resolved and semi-resolved young star clusters ($\lesssim$30\,Myr).

Furthermore, we demonstrate that FGC-based structural parameters for clusters of log$\rm{(M/M_{\sun})}\lesssim 3.0$ and log$\rm (t/yr)\lesssim 8.0$ are rather unconstrained and show large uncertainties due to stochastic effects inherent to low-mass star clusters. Hence, in the case of such objects, additional constraining criteria are necessary to study the cluster size and structure evolution reliably.

Finally, we show that the randomness of low-mass star cluster projections towards the line of sight introduces significant uncertainties (up to 0.1\,mag for CIs) in aperture photometry measurements and typically $\sim$10-20\% uncertainties for structural parameters, depending on the interplay between cluster physical and radial profile parameters with aperture sizes. Therefore, this effect should be accounted for when estimating the total error budget of aperture photometry measurements.

We emphasise that none of the experiments performed in this study consider sky background effects. Thus, all distortions and errors of star cluster photometric and structural measurements, as well as derived physical cluster parameters, discussed in the paper, represent the best achievable accuracy using aperture photometry methods. The inclusion of any other precision limiting factors, such as the sky background or resolved bright field stars, would increase reported uncertainties. Also, it is important to mention that systematic problems, such as parameter degeneracies \citep{Worthey1994, Bridzius2008, Whitmore2023} or differences between stellar evolution and population synthesis models, can in some cases have a greater impact on our ability to determine accurate star cluster parameters compared to random errors due to stochastic effects.

\begin{acknowledgements}
We are grateful to the anonymous referee for many constructive suggestions, which improved the paper. This research has made use of: the NASA/IPAC Extragalactic Database (NED), which is funded by the National Aeronautics and Space Administration and operated by the California Institute of Technology; SAOImage DS9, developed by the Smithsonian Astrophysical Observatory; Astropy (\url{http://www.astropy.org}), a community-developed core $\tt Python$ package for Astronomy \citep{Astropy2013, Astropy2018}; $\tt APLpy$, an open-source plotting package for $\tt Python$ \citep{aplpy2012}; SciPy, an open-source scientific library for $\tt Python$ \citep{scipy2020}; Matplotlib, a plotting library for $\tt Python$ \citep{Hunter2007}. Computations were performed on the supercomputer GALAX of the Center for Physical Sciences and Technology, Lithuania. 
\end{acknowledgements}

\bibliographystyle{aa}
\bibliography{49680corr}

\end{document}